# Improving the State of the Art for Training Human-AI Teams

**Technical Report #5**

# Individual Differences and Team Qualities to Measure in a Human-AI Teaming Testbed

**February 2024**

*Lillian Asiala and Jim McCarthy*

*Sonalysts, Inc.*

This Page Intentionally Blank

# Table of Contents

# List of Tables

# Acronyms

# Abstract

Sonalysts, Inc. (Sonalysts) is working on an initiative to expand our expertise in teaming to include Human-Artificial Intelligence (AI) teams. The first step of this process is to develop a Synthetic Task Environment (STE) to support our original research. Prior knowledge elicitation efforts within the Human-AI teaming research stakeholder community revealed a desire to support data collection using pre- and post-performance surveys. In this technical report, we review a number of constructs that capture meaningful individual differences and teaming qualities. Additionally, we explore methods of measuring those constructs within the STE.

This Page Intentionally Blank

# 1 BACKGROUND

A consensus report produced for the Air Force Research Lab by the National Academies of Sciences, Engineering, and Mathematics documented a prevalent and increasing desire to support Human-AI teaming across military service branches (NASEM, 2021). Sonalysts, Inc. (Sonalysts) has begun an internal initiative to explore ways to maximize the performance of Human-AI teams. To provide a foundation for our research, Sonalysts is developing a Concept of Operations (CONOPS) for a Synthetic Task Environment (STE) that can serve as a testbed. As a part of that CONOPS, the research team wants to include measures of individual differences and team qualities that we anticipate will have meaningful impacts on performance. This report will review the various constructs that we have identified for inclusion in the STE, and methods for conducting and validating those measurements.

Two distinct categories of constructs have a demonstrable impact on team performance. The first is the *individual trait group*, which is focused on relatively stable characteristics that arise from an individual's personality or as a function of their developmental background within a particular environment. These traits can be described broadly in terms of temperament and ability. Traits related to temperament include those that fall within traditional models of personality, such as those belonging to the "Big Five" model of personality: extraversion, agreeableness, emotional stability, conscientiousness, and openness to experience. They also encompass the person's general predisposition to individualism or collectivism, interest in working in groups, and Locus of Control. We also consider general abilities such as cognitive and interpersonal abilities (*e.g.,* emotional competence and intelligence) to fall under this "trait level" category. Generally, these qualities affect many aspects of an individual's experience within and beyond the team dynamic.

The second group of constructs is specific to *working in a team environment*, and has to do with the beliefs and attitudes an individual has about their specific team. Teamwork-related beliefs and attitudes are more likely to be informed by an individual's experience operating in *particular* group settings, and therefore could be more amenable to change over time. While these constructs are conceptualized as an individual's attitude about a team, the measures for these constructs tend to ask respondents to reference their experience with a *particular* team at the time of measurement (as opposed to their attitude toward teams generally, or their proclivity to work well with others). An individual's experience or training could conceivably change these constructs.

Both of the groups of constructs described above are believed to influence team performance on the individual level. However, a subset of these constructs has been examined in terms of how the constructs contribute to team composition, leading to both an individual measurement level of analysis and a team level of analysis. Each of the following sections features a discussion of constructs in terms of how they have been shown to influence team performance at the individual and team level.

This Page Intentionally Blank

## 2 TRAIT-LEVEL CONSTRUCTS

Trait-level constructs are aspects of an individual's personality or ability that they bring to the operational environment. These characteristics are not thought to be changeable through experience or training. In this section, we will discuss the trait-level constructs that research has shown to impact team performance or an individual's perception of the team.

### 2.1 Big Five Personality Traits

Numerous studies have found connections among particular personality traits, perceptions of the team, and overall team performance. The five factor model – extraversion, agreeableness, conscientiousness, neuroticism (also called emotional stability), and openness to experience – has provided a useful measurement framework for this kind of research (Peeters et al., 2006a). In this section, we will discuss the Big Five model of personality, operationalization and measurement approaches, and the relationship between operationalization and measurement in studies on teaming.

The Big Five traits are supported by decades of research across many populations and cultures to describe universal personality dimensions that influence affect, behaviors, and cognitions (Zillig et al., 2002). This framework is lexically represented, meaning that it is frequently measured by assessing level of agreement with adjectives associated with each of the personality factors (Caprara et al., 1993). Roccas et al., (2002) provide the following useful descriptions of each dimension:

> **Extraversion**: A high score on this dimension indicates high sociability, assertiveness, talkativeness, and generally higher activity levels. Alternatively, a lower score indicates someone is more reserved and cautious. Extraversion tends to be a personality trait compatible with an achievement orientation (particularly if that achievement requires socializing productively) and novelty-seeking.

> **Agreeableness**: High agreeableness is associated with a good-natured, compliant bearing. Individuals high on this dimension are typically modest, gentle, and cooperative, while those who score lower on this dimension tend toward irritability, ruthlessness, suspicion, and inflexibility. Agreeableness is compatible with concern for the welfare of others, not violating norms or upsetting others, and compliance with cultural and religious norms. It is incompatible with the desire to dominate others.

> **Openness to Experience**: Individuals with a high score on this dimension tend to be intellectual, imaginative, sensitive, and open-minded. A low score indicates the individual is more down-to-earth, insensitive, and conventional. Openness tends to be compatible with goals related to self-directedness, tolerance, and novelty, while conflicting with goals related to conformity and security.

> **Conscientiousness**: A high score on this dimension indicates a high level of care, thoroughness, organization, and scrupulousness, and a deep sense of responsibility. Someone low on this dimension may be somewhat irresponsible, disorganized, and unscrupulous. Two aspects of conscientiousness are a proactive will to achieve a goal, and the inhibition of impulsive behaviors. It is compatible with goals related to both achievement and conformity.

> **Neuroticism**: A high score on this dimension tends to be related to feelings of anxiousness, depression, anger, and insecurity, whereas a lower score is indicative of a calm, poised, and emotionally stable temperament.

### 2.1.1 Approaches to Operationalizing Personality Dimensions in Teams

Researchers have proposed two approaches for operationalizing personality dimensions in teaming research. This first is to look at traits at the level of the individual. The second is to aggregate personality variables at the level of the team. In both cases, the researchers then relate personality measures to team performance variables (Peeters et al., 2006a).

Measuring personality traits at the level of the individual is straightforward. Within this approach, researchers measure an individual's scores on various dimensions and examine how those scores affect their perception of and performance within that team.

Operationalizing personality variables at the team level is more complicated, and researchers accomplish it using several methods. Some researchers operationalize personality variables at the team level using a method known as "Team Personality Elevation" (TPE). TPE refers to a team's *average* level of a particular personality dimension. For instance, a high TPE on a particular personality dimension indicates that at least one team member has a high score on that dimension – enough to "elevate" the team's average score. The idea behind this approach is that the amount of that personality dimension across team members is directly related to the outcome variable.

Another method of operationalizing personality dimensions in teaming is to examine how members of a team differ from one another on a particular personality dimension. This method is referred to as "Team Personality Diversity" (TPD). In this method, researchers operationalize the team's personality as the *variance* of a particular personality trait represented within a team. This variance is believed to have an effect on the outcome variable of interest. That is, no matter how much of the dimension is represented in the team, it is the variability of the dimension among team members that influences the outcome.

Researchers have also operationalized team personality dimensions as the lowest team member score. For example, Neuman and Wright (1999) measured personality dimensions at both the individual and team level, and used the lowest individual score on the dimension to represent the team. The belief behind this method is that the lowest score indicates the limitations of the team, which are more indicative of success than the highest scoring teammate. Thus, the lowest score is the best indication of expected performance (Neuman & Wright, 1999).

There are reasons to think that different operationalization methods would lead to different findings regarding the relationship between personality dimensions and aspects of team performance. For example, greater TPD on personality dimensions might be particularly useful if the team needs to conduct a diverse set of tasks, and roles are differentiated significantly (assuming that team members are aligned to the best fitting role for them). On the other hand, overall TPE on certain dimensions might make a difference if the team is relatively "flat" (*i.e.,* responsibilities are handled on a relatively egalitarian basis, instead of hierarchically). In that case, a higher average level of a particular trait (such as conscientiousness or extraversion) among team members may yield better results than a wide range of scores on this dimension.

### 2.1.2 Measuring the Big Five

The operationalization methods outlined in section 2.1.1 are all achieved via individual measures of each personality dimension, and by using the scores according to the desired method. This means that regardless of whether evaluating minimum, maximum, TPE, or TPD for a particular characteristic and team, the construct is first captured at the individual level; through a validated measure of the Big Five.

There are several validated measures of Big Five personality traits available for inclusion in the CONOPS. A popular measure is the Revised NEO Personality Inventory (NEO-PI-R). However, this measure is not available in the public domain. Another is a measure based on the International Personality Item Pool (IPIP), such as the IPIP-NEO-120 (Johnson, 2014) and the mini IPIP (a 20-item subset of the original 50 items, with good validity and reliability; Donnellan et al., 2006). The suitability of these measures for inclusion in the testbed CONOPS is summarized below in Table 1.

Table 1. Summary of Big Five Personality Measures

| Personality Measure | Qualities | Suitability for Inclusion in CONOPS |
|---|---|---|
| ***Mini IPIP***<br><br>Donnellan, M. B., Oswald, F. L., Baird, B. M., & Lucas, R. E. (2006). The mini-IPIP scales: tiny-yet-effective measures of the Big Five factors of personality. *Psychological assessment*, *18*(2), 192. | Availability: Publically Available<br><br>Reliability and Validity: Validated and used across multiple psychological areas of research.<br><br>Report Type: Self<br><br>Applicability to the Task Environment: This measure is fairly universally applicable and reasonable for the task environment.<br><br>Applicability to the Target Participant: This measure is fairly universally applicable and reasonable for the audience. | **Suitable.** |
| ***IPIP-NEO-120***<br><br>Johnson, J. A. (2014). Measuring thirty facets of the Five Factor Model with a 120-item public domain inventory: Development of the IPIP-NEO-120. *Journal of research in personality*, *51*, 78-89. | Availability: Publically available<br><br>Reliability and Validity: Validated and used across multiple psychological areas of research (see Johnson, 2014).<br><br>Report Type: self-report<br><br>Applicability to the Task Environment: This measure is fairly universally applicable and reasonable for the task environment.<br><br>Applicability to the Target Participant: This measure is fairly universally applicable and reasonable for the audience. However, it is a longer measure involving 120 items and may be cumbersome to involve in research. | **Suitable (but may be long).** |

## 2.2   Personality Dimensions and Teaming Performance

As an illustration of the operationalization methods introduced in Section 2.1.1, Neuman, Wagner and Christiansen (1999) found that traits of conscientiousness, agreeableness, and openness to experience were positively related to team performance through TPE (*i.e.,* a higher overall mean for the team on those dimensions was positively related to team performance). Extraversion and emotional stability, on the other hand, were positively related to performance through TPD (indicating the variance among team members on these traits affected team performance).

A meta-analysis led by Peeters and colleagues (2006b) also sought to investigate how team composition affected team performance. Like Neuman et al. (1999), their study found the elevation of agreeableness in teams (TPE) had a positive relationship with team performance, while variability (TPD) in that dimension had a negative effect on performance. The same pattern was true for conscientiousness. Thus, uniformly high levels of conscientiousness and agreeableness improved team performance. Peeters et al (2006b) also found that relationships between personality and performance appeared to be a function of the type of team under investigation. For instance, they found that universal conscientiousness affected team performance in a professional setting, but not in an academic setting (*i.e.,* student teams). Conversely, they found an effect of emotional stability in student teams, but not professional teams. Professional teams overall seemed to perform better when the average conscientiousness and agreeableness of teammates was higher, whereas trait emotional stability seemed to be more important for the performance of student teams. In terms of diversity, professional and student teams also differed regarding the effect of variability of emotional stability and openness on team performance. Student teams performed better if they were more homogenous on emotional stability, whereas more similarity on openness to experience was more important for the performance of professional teams (Peeters et al., 2006).

Peeters and another set of authors (2006a) focused their investigation on TPD and team satisfaction. Peeters et al. (2006a) looked at the relationships among an individual's satisfaction with the team and the team members' Big Five personality traits. Specifically, an individual's satisfaction with the team was regressed on an individual score, a dissimilarity (to the team) score, and an interaction score. The central question for their study was to assess how team composition, in terms of personalities of the team members, influenced the satisfaction individual team members had within that particular team. The authors defined team satisfaction as encompassing dimensions related to team members, work processes, and the team's viability. The researchers measured team satisfaction with a three-item questionnaire, where all three items loaded onto the same factor in a Confirmatory Factor Analysis (CFA). Overall, agreeableness TPD and emotional stability TPD positively predicted team satisfaction. Conscientiousness TPD was a negative predictor for team satisfaction, so the more dissimilar team members were from their teammates on conscientiousness, the less satisfied with their team they tended to be. There was also a negative effect of dissimilarity in extroversion, but only for those who scored low on extroversion. These findings are particularly interesting in the discussion of task-related and team-related personality traits. Conscientiousness is a task-related trait, while agreeableness, stability, and extroversion are "team-related." The "team related" traits have a direct positive effect on team satisfaction via TPE, while conscientiousness affected performance through TPD.

## 2.3 Emotional Intelligence

Emotional intelligence is a construct that has been widely studied within educational environments and the workplace. While it has perhaps not had direct connections within teaming research specifically, it has inherent application to teams of people who work toward common goals in those contexts. Mayer and Salovey (1997) developed one of the most frequently cited definitions of emotional intelligence, describing it as "…the ability to perceive accurately, appraise, and express emotion; the ability to understand emotions and emotional knowledge; and the ability to regulate emotions to promote emotional and intellectual growth." In the following sections, we will explore the ways that emotional intelligence has been operationalized and measured, the suitability of various measures of emotional intelligence within the TCT, and human/autonomy teaming research more broadly.

### 2.3.1 *Operationalizing Emotional Intelligence*

Emotional Intelligence (EI) can be conceptualized as an "ability," a "trait," or a combination of the two (*i.e.,* measured with a "mixed" measure; Bru-Luna et al., 2021; Butler et al., 2022). An "ability" model of EI involves viewing it as a skill made up of several capacities the individuals have in managing their emotions and the emotions of others, including the perception, evaluation, and expression of emotions; emotional facilitation of thought; understanding and analysis of emotions; and reflective regulation of emotions (Bru-Luna et al., 2021). Typically, researchers conceptualize ability model measures of EI as performance measures.

A popular ability-based measure of EI is the Mayer-Salovey-Caruso Emotional Intelligence Test (MSCEIT) (Mayer, 2002). Psychometric evaluation of the MSCEIT has shown evidence for a factor structure consistent with the following four dimensions:

- Emotional perception (the identification of emotion in faces, music, and stories)
- Emotional integration (relating emotions to other mental sensations and using emotion in reasoning and problem-solving)
- Emotional understanding (solving problems such as knowing which emotions are similar or opposite)
- Emotional management (understanding the implication of social acts on emotion and the regulation of emotion in the self and others)

Another popular ability-based measure of EI is Wong and Law's Emotional Intelligence Scale (WLEIS; Wong & Law, 2002). This measure relies on the four-component model of EI established by Mayer and Slovey (1997). Wong and Law originally developed the measure for use in China, but it has demonstrated reasonable consistency in other cultures (Libbrecht et al., 2014). The Genos Emotional Intelligence Inventory (Genos EI) was created to develop employees in the workplace, and is an ability-oriented "behavior-based" assessment of workplace behavior. Rather than being a measure of EI outright, the Genos EI is a measure of an individual's ability to demonstrate workplace behaviors that represent EI (Palmer et al., 2009). The dimensions measured by Genos EI include emotional self-awareness, emotional expression, emotional awareness of others, emotional reasoning, emotional self-management, emotional management of others, and emotional self-control. An important note about this measure is that a central goal in its development was workplace face validity, rather than having a theoretical foundation. Potentially, this limits its utility outside the corporate work environment. It is available in various lengths and in self-report or rater-report format.

The "trait" model of EI considers it as a persistent behavior pattern over time (Bru-Luna et al., 2021). An example of a trait-based measure of EI is the Trait Emotional Intelligence Questionnaire (TEIQue), which is focused on an individual's "emotional self-efficacy." The TEIQue measures 15 facets including adaptability, assertiveness, emotional expression, emotional management, emotional perception, emotion regulation, and others. It has demonstrated construct validity with other similar measures (Petrides, 2009).

Another trait measure is the Trait Meta-Mood Scale (TMMS), which has been described as a self-report measure assessing the extent to which an individual attends to moods, experiences clarity of their own experiences of moods, and attempts to repair mood states (Salovey et al., 1995). A third trait-based measure of EI is based on Mayer and Salovey's (1995) model of EI, and goes by several names including:

- The Assessing Emotions Scale (AES)
- The Emotional Intelligence Scale, the Self Report Emotional Intelligence Test (SREIT)
- Schutte Self-Report Emotional Intelligence Test (SSEIT)

- Schutte Emotional Intelligence Scale
- Schutte Self-Report Inventory (SSRI)

The TMMS and SSRI measures do not appear to be publically available.

The most popular way of conceptualizing EI is to use a "mixed-model" concept. The mixed-model approach assesses multiple dimensions of EI (Butler et al., 2022). Mixed-model assessments of emotional competency combine measures of multiple social-EI qualities that include personality, motivation, emotionally based skills, and other areas (Butler et al., 2022). The mixed-model generally conceptualizes the construct as a blend of traits, competencies, and abilities (Bru-Luna et al., 2021). For example, one conceptualization is that EI comprises intrapersonal skills, interpersonal skills, adaptation skills, stress management skills, and general mood (Bar-On, 2003). Another involves the recognition of one's own emotions, management of emotions, self-motivation, recognition of emotions in others, and management of relationships (Goleman et al., 2002). Mixed-model measures are typically self-report (as opposed to demonstrations of knowledge), which leaves them vulnerable to bias. However, they can also include ratings from individuals who know the target of measurement well (such as supervisors and peers), which can correct for this bias.

Butler et al. (2022) provide a comprehensive breakdown of many of these mixed model measures. They include:

- The Emotional Quotient Inventory 2.0 (EQ-i 2.0): A mixed-model measuring social and emotional traits and their influence on an individual's well-being. There is also a multi-rater version of the instrument that combines self-ratings and ratings of those who know the individual.
- The Social and Emotional Development Inventory (SED-I): A student-focused model designed to improve social/emotional development of students in higher education. The scores can be used to create learning plans for students based on their areas of competence.
- The Personal-Interpersonal Competence Assessment (PICA): A revision of the student-focused SED-I measure
- The Emotional Intelligence Appraisal Assessment: A measure intended to increase EI through specific strategies.
- The Emotional Competence Inventory (ECI): A measure designed to assess the emotional competencies of individuals and organizations. It measures 18 competency areas through ratings of others.
- The Emotional and Social Competency Inventory (ESCI): A measure of a subset of 12 competency areas of the 18 measured by the ECI through self-scores and ratings of others.
- The Emotionally Intelligent Leadership for Students (EILS) Inventory: A self-report assessment of emotionally intelligent leadership within a student environment, measuring consciousness of context, consciousness of self, and consciousness of others.

Some researchers expand the construct from EI to a more comprehensive Emotional Competence (EC) construct. Take for instance, Goleman, Boyatzis, and McKee's (2002) Emotional Competency Inventory (ECI). This is a self-assessment with the following dimensions:

- Self-awareness (emotional self-awareness, accurate self-assessment, self-confidence)
- Self-management (emotional self-control, trustworthiness, conscientiousness, adaptability, optimism, achievement orientation, and initiative)
- Social awareness (empathy, organizational awareness, and service orientation)
- Relationship management (developing others, inspirational leadership, influence, communication, change catalyst, conflict management, building bonds and teamwork, and collaboration).

Note that the dimensions listed above include both emotional abilities (like social awareness) and the product of those abilities (like relationship management) compared to the ability-focused dimensions of EI. An appealing feature of the EC construct is that it integrates skills and qualities that are known to affect performance (Offerman & Riley, 2004). Offerman and Riley (2004) assessed the impact of EC on team performance and team assessment. Particularly, they contrasted the effects of EC with Cognitive Ability (CA) on performance. The authors expected cognitive ability to predict better individual academic performance, and emotional competency to predict better team performance (though both were expected to have a positive relationship with individual and team-related performance outcomes). Further, these authors predicted that the collective EC representation within the team would predict team performance. They predicted that both individuals and teams with higher levels of emotional competency would have more positive attitudes toward their team experience than those with lower emotional competency. They also predicted that those individuals with higher emotional competency on a team would be more likely to be identified as leaders and be rated higher on leadership effectiveness than those with low emotional competency. Participants for the study were randomly assigned project teams for a business administration course in human relations. They used the ECI-U to measure emotional competency, which is a version of the ECI modified for university students. In addition, authors measured cognitive ability and personality using a measure of the Big Five personality traits. Measures of academic and team achievement included student exam grades, team projects, team attitudes, and leadership emergence (assessed through peer rankings) and effectiveness (measured by ratings).

As predicted, cognitive ability was a predictor for individual performance scores (emotional competency was not), and emotional competency (particularly self-management, social awareness, and relationship management sub-scores) were more effective at predicting team project performance than cognitive ability. Additionally, total team EC score was positively associated with team project scores. Average EC also appeared to predict team attitudes, as well as the emotional competency of the lowest scoring team member. At the individual level, the relationship management subscale accounted for significant variance in the positive relationship with team attitudes. At the group level, the same findings held (although not to a significant degree). Both leadership variables were significantly related to EC total scores and sub-scale measures of self-awareness and relationship management.

The ECI would be a good measure to consider incorporating into the testbed; however, a free version of the instrument does not appear to be available[1].

### 2.3.2 Measuring Emotional Intelligence

Overall, the suitability for inclusion of measures of EI (or EC) in the testbed involves an evaluation of the measure's availability, applicability to the population most likely to use the testbed, applicability to the task environment, and psychometric qualities (such as theoretical underpinnings, validity, and reliability). Table 2 considers the suitability of each of these measures for inclusion in the testbed CONOPS on the basis of these factors.

---

[1] Information about the ECI in its various forms is available at the following hyperlink: https://www.danielgoleman.info/ei-assessments/.

Table 2: Summary of Emotional Intelligence Measures

| Emotional Competency Measure | Qualities | Suitability for Inclusion in CONOPS |
|---|---|---|
| *Trait Emotional Intelligence Questionnaire (TEIQue)*<br><br>Petrides, K. V. (2009). Psychometric properties of the trait EI questionnaire (TEIQue). In *Assessing EI: Theory, research, and applications* (pp. 85-101). Boston, MA: Springer US. | Availability: Publically Available<br><br>Operationalization Variant: Trait measure<br><br>Reliability and Validity: Good<br><br>Report type: Self-report<br><br>Applicability to the Task Environment: The TEIQue-SF in particular has been amended to measure "task-related stress," which is especially applicable to the kind of stress we anticipate will influence team members in the TCT, making it a good fit for our purposes.<br><br>Applicability to the Target Participant: The questions are designed to be generalizable across many populations and teaming/working conditions, making this a good fit for the TCT. | **Suitable**: Overall, the TEIQue (particularly its short form) seems to be a good fit for inclusion in the TCT. The only drawback to this measure is that it is a trait-level rather than mixed measure, which would be ideal. |

| Emotional Competency Measure | Qualities | Suitability for Inclusion in CONOPS |
|---|---|---|
| ***Genos Emotional Intelligence Inventory (Genos EI)***<br><br>Palmer, B. R., Stough, C., Harmer, R., & Gignac, G. (2009). The Genos Emotional Intelligence Inventory: A measure designed specifically for workplace applications. *Assessing EI: Theory, research, and applications*, 103-117. | Availability: Publically Available<br><br>Operationalization Variant: Ability<br><br>Reliability and Validity: This measure had reasonably good reliability but did not have a theoretical relationship to the construct of EI.<br><br>Report Type: Multi-rater measure<br><br>Applicability to the Task Environment: This is a measure of certain workplace behaviors that have no known relationship to the task environment presented in the TCT. Therefore, the predictive quality of this measure to performance in the TCT is unknown.<br><br>Applicability to the Target Participant: This measure is aimed at participants working in a standard corporate environment, not the operational teaming environment presented in the TCT. | **Suitable, not recommended**: This measure has good psychometric properties and several versions available. However, it was not intended to capture the construct of EI directly, but is a measure of certain workplace behaviors that have an unknown relationship to the target operational environment for the testbed. It is also a multi-rater tool which is not ideal for inclusion in the TCT. |

| Emotional Competency Measure | Qualities | Suitability for Inclusion in CONOPS |
|---|---|---|
| ***The Social and Emotional Development Inventory (SED-I)***<br><br>Seal, C. R., Beauchamp, K. L., Miguel, K., & Scott, A. N. (2011). Development of a self-report instrument to assess social and emotional development. *Journal of Psychological Issues in Organizational Culture*, *2*(2), 82-95. | Availability: Publically Available<br><br>Operationalization Variant: Mixed<br><br>Reliability and Validity: The SED-I has good internal consistency, and moderate consistency between hypothesized factors and factor structure.<br><br>Report Type: Self-report<br><br>Applicability to the Task Environment: The SED-I is tailored to a higher education environment and is designed to assist in the development of student learning plans. It is not designed to address the challenges associated with the operational task presented by the TCT.<br><br>Applicability to the Target Participant: SED-I is aimed at students in higher education, making it a poor fit for the research participants we anticipate will use the TCT. | **Suitable, not recommended:** Significant aspects of this measure are promising for research purposes. However, its purpose is to measure emotional development of students in higher education and assist in the creation of learning plans to support students, which is not consistent with the goals of the TCT. |

| Emotional Competency Measure | Qualities | Suitability for Inclusion in CONOPS |
|---|---|---|
| *The Personal-Interpersonal Competence Assessment (PICA)*<br><br>Seal, C. R., Miguel, K., Alzamil, A., Naumann, S. E., Royce-Davis, J., & Drost, D. (2015). Personal-interpersonal competence assessment: A self-report instrument for student development. *Research in Higher Education Journal*, 27. | Availability: Publically Available<br><br>Operationalization Variant: Mixed<br><br>Reliability and Validity: This measure had good internal consistency and support for the hypothesized four-factor structure.<br><br>Report Type: Self-report<br><br>Applicability to the Task Environment: The PICA is a shortened form of the SED-I, which is designed to assist in the development of student learning plans. It is not designed to address the challenges that would be associated with the operational task presented by the TCT.<br><br>Applicability to the Target Participant: As with the SED-I, the PICA is aimed at students in higher education, making it a poor fit for the research participants we anticipate will use the TCT. | **Suitable, not recommended:** The PICA is derived from the SED-I. It is intended for the same population – students of higher education. The goal of the revision was to improve the student professional development aspect of the measure. Therefore, it has the same issues as SED-I. |

| Emotional Competency Measure | Qualities | Suitability for Inclusion in CONOPS |
|---|---|---|
| *The Emotional Quotient Inventory 2.0 (EQ-i 2.0)*<br><br>Bar-On, R. (2003). BarOn Emotional Quotient-Inventory (BarOn EQ-i®). *EI Consortium*, *1*(8). | Availability: Not publically available<br><br>Operationalization Variant: Mixed<br><br>Reliability and Validity: This measure shows strong reliability (internal consistency and test/retest).<br><br>Report Type: Self Report<br><br>Applicability to the Task Environment: The BarOn EQ-i is designed for clinical, educational, forensic, medical, corporate, human resources, and research settings, and is likely applicable to human/autonomy teaming research as well.<br><br>Applicability to the Target Participant: This measure is aimed at measuring EI in many populations of adults; however, it is a long measure (133 items) which, in combination with other desired measures, may be imposing and take a lot of time to for participants to complete. | **Unsuitable:** This is a promising measure for research purposes, but is not considered because it is not publically available. |

| Emotional Competency Measure | Qualities | Suitability for Inclusion in CONOPS |
|---|---|---|
| **The Emotional Intelligence Appraisal Assessment**<br><br>Butler, L., Park, S. K., Vyas, D., Cole, J. D., Haney, J. S., Marrs, J. C., & Williams, E. (2022). Evidence and strategies for including emotional intelligence in pharmacy education. *American Journal of Pharmaceutical Education*, *86*(10), 1095-1113. | Availability: Not publically available<br><br>Operationalization Variant: Mixed<br><br>Reliability and Validity: This measure demonstrates strong reliability. It is unclear what its theoretical underpinnings are.<br><br>Report Type: Self-report<br><br>Applicability to the Task Environment: From the available descriptions of this measure, the EI Appraisal Assessment appears to be targeted toward general use in the workplace. It is not specific to the operations presented by TCT, but does not necessarily run counter to them. However, it may fail to address the specific emotional challenges associated with teaming environments under pressure.<br><br>Applicability to the Target Participant: This measure seems mainly applicable to those in a corporate environment, but is probably applicable to many domains. | **Unsuitable:** This measure is not publically available. |

| Emotional Competency Measure | Qualities | Suitability for Inclusion in CONOPS |
|---|---|---|
| *The Emotional Competence Inventory (ECI)*<br><br>Goleman, D., Boyatzis, R., & McKee, A. (2002). *Primal leadership: Realizing the power of emotional intelligence.* Boston: Harvard University Press. | Availability: Not publically available<br><br>Operationalization Variant: Mixed<br><br>Reliability and Validity: Reliability of the multi-rater assessment has shown to be strong.<br><br>Report Type: Multi-rater assessment (there is a self-report assessment, but its psychometric qualities are worse than the multi-rater version).<br><br>Applicability to the Task Environment: This measure is designed to capture day-to-day work behaviors, which may not be an ideal fit with the types of behaviors required for performance in the TCT operational environment.<br><br>Applicability to the Target Participant: This measure is strongest in its multi-rater form. Therefore, it is not useful for teams who may be placed together for the first time in a research setting and do not know one another well. | **Unsuitable:** This measure is not publically available. Additionally, this measure is strongest in its multi-rater form, which may not be realistic for incorporating in the TCT, as it requires respondents to know the target of evaluation well. |

| Emotional Competency Measure | Qualities | Suitability for Inclusion in CONOPS |
|---|---|---|
| *The Emotional and Social Competency Inventory (ESCI)*<br><br>Butler L, Park SK, Vyas D, Cole JD, Haney JS, Marrs JC, Williams E. Evidence and Strategies for Including Emotional Intelligence in Pharmacy Education. Am J Pharm Educ. 2022 Dec;86(10):ajpe8674. doi: 10.5688/ajpe8674. Epub 2021 Oct 25. PMID: 34697015; PMCID: PMC10159398. | Availability: Not publically available<br><br>Operationalization Variant: Mixed<br><br>Reliability and Validity: This measure appears to have strong reliability.<br><br>Report Type: Multi-rater<br><br>Applicability to the Task Environment: This measure is a subset of the ECI and therefore is aimed to capture day-to-day work behaviors, which may not be an ideal fit for research within the TCT.<br><br>Applicability to the Target Participant: Like the ECI, this is a multi-rater measure which requires that participants know one another well. | **Unsuitable:** This measure is not publically available. This measure is a sub-set of the ECI. |

| Emotional Competency Measure | Qualities | Suitability for Inclusion in CONOPS |
|---|---|---|
| *The Emotionally Intelligent Leadership for Students (EILS) Inventory*<br><br>Butler L, Park SK, Vyas D, Cole JD, Haney JS, Marrs JC, Williams E. Evidence and Strategies for Including Emotional Intelligence in Pharmacy Education. Am J Pharm Educ. 2022 Dec;86(10):ajpe8674. doi: 10.5688/ajpe8674. Epub 2021 Oct 25. PMID: 34697015; PMCID: PMC10159398. | Availability: Not publically available<br><br>Operationalization Variant: Mixed<br><br>Reliability and Validity: This measure has strong reliability.<br><br>Report Type: Self-report<br><br>Applicability to the Task Environment: This measure is aimed at students in a learning environment, and is not geared toward the type of operational performance presented by the TCT.<br><br>Applicability to the Target Participant: This measure is aimed at students in high school, college, and those pursuing graduate studies, and the work they perform in those environments. Therefore, it is likely not a good fit for potential participants in the TCT. | **Unsuitable:** This measure is not publically available. It is also aimed specifically at a student population. |

## 2.4 Individualist and Collectivist Orientation

Individualism and collectivism are opposing ends of a singular dimension of culture that has been well studied and documented within the psychological literature (Triandis & Gelfand, 2012). It is best described as the nature of the relationship between an individual and group (Triandis & Gelfand, 2012). Triandis' work began as a collaboration between the Office of Naval Research and the University of Illinois. The goal was to enable Naval officers to manage cultural differences as a part of their work. The program of research began as a qualitative look at subjective aspects of culture (such as beliefs, attitudes, expectations, and norms). Eventually, coherent themes across cultures began to emerge, one of which was the theme of individualism and collectivism (although the construct was actually coined by Geert Hofstede performing a similar analysis in parallel to Triandis' work) (Triandis & Gelfand, 2012). According to Hofstede's definition, *individualism* pertains to cultures in which ties among individuals are loose, and an individual's primary responsibility is to oneself and immediate family. Collectivism refers to a culture in which people are born into, integrated within, and protected by cohesive in-groups to which they are expected to be completely loyal. From these initial definitions, subsequent research was conducted to further understand and refine these constructs (Triandis & Gelfand, 2012).

## 2.4.1 Operationalizing Individualism and Collectivism

This cultural dimension is conceptualized in the literature as a "value" – a guide for actions, behaviors, and attitudes within a given situation (Bell, 2007). Research on developing measurements of individualism and collectivism began with work in the 1980s, and found psychometric evidence for these constructs in a cross-cultural research study spanning nine countries. The culture-level analysis yielded two factors associated with individualism – self-reliance and hedonism, and two factors associated with collectivism – family integrity and interdependence with society. As the concepts evolved, other ways of conceptualizing this cultural dimension included whether a person puts an emphasis on autonomy or being embedded in groups, or whether an emphasis is placed on independence or interdependence. By the end of the decade, Triandis published a theory that placed individualism and collectivism in the context of how frequently different facets of the self (private, public, or collective) were sampled in a social context. Members of an individualistic culture were more likely to call upon the "private self" in a social setting than the "collective self," and vice versa for members of collectivist cultures (See Triandis, 1989).

In their 2012 paper, Triandis and Gelfand summarized a number of attributes of individualism and collectivism (a representative selection of these attributes is listed below)

| Individualism | Collectivism |
|---|---|
| - The self is independent of the group. | - The self is interdependent with the group. |
| - Individual goals take precedence over the group's goals. | - Group goals are prioritized over individual goals. |
| - Group and individual goals may be different. | - Group and individual goals are more likely to be the same. |
| - Attitudes, personal needs, individual rights, and contracts established between an individual and others take precedence. | - Norms, obligations, and duties guide behavior. |
| - Individuals are more likely to leave unsatisfactory relationships. | - Individuals are more likely to stay in unpleasant groups or relationships. |
| - The focus of attention is on the individual, and relationships are in the background. | - The focus of attention is on relationships, and individuals are in the background. |
| - Determinants of behavior tend to be internal. | - Determinants of behavior tend to be external. |
| - Individuals tend to see themselves as invariant and the environment as change-able. | - Individuals tend to view the environment and themselves as change-able. |
| - Self-efficacy is enhanced when working alone. | - Self-efficacy is enhanced when working in groups. |
| - General "resistance" to teams. | - Perception of teams as "entities" with dispositions of their own. |

- Emphasis on high task orientation and low socioemotional behaviors as important for success.
- Lesser tendency to hold groups and organizations responsible for failed actions.
- Less likely to endorse charismatic, team-oriented leadership.
- Less tolerant of "talking behind a subordinate's back," rather than direct communication.
- Conflict is perceived to be about violations of individual rights and autonomy.

- Emphasis on socioemotional behaviors as important for group success.
- Greater tendency to hold groups and organizations responsible for failed actions.
- More likely to endorse charismatic, team-oriented leadership.
- Appreciative of "talking behind a subordinate's back," rather than direct communication.
- Conflict is perceived to be about violations of duties and obligations.

A few of the attributes listed above may be particularly salient to human/autonomous teaming research, especially to the perception one has of the role of autonomous agents on a team. For example, the way each culture views relationships and the obligations an individual has to the relationships they maintain vary starkly between individualist and collectivist cultures. Would the obligation that an individual with a collectivist orientation feels toward a human teammate extend to an synthetic member of the team? Would the individualist be more likely to accept AI agents as teammates than a collectivist, given the individualist's emphasis on task orientation, and relative disregard of socioemotional behaviors? How would an individual's mental representation of "team" from either cultural background affect their perception of AI agents? Would the norm for teams within collectivist cultures to take on a unique disposition further enhance or detract from the perception of the AI agents?

Another notable finding from the literature that Triandis and Gelfand (2012) discuss are the situational conditions that give rise to individualism or collectivism in the first place. Conditions that threaten the in-group tend to increase the probability of collectivism, whereas conditions that are more likely to reward individual actions are more likely to lead to individualism. As a concrete example of this phenomenon, tasks that a single group member can complete are more likely to promote individualism, while tasks that are high in interdependence promote collectivism. Research even demonstrates that subtle priming of collectivism or individualism can even elicit responses in line with one of those orientations (Triandis & Gelfand, 2012).

### 2.4.2 Individualism and Collectivism in Teams

While much of the discussion of individualism and collectivism up to this point is described at the level of an entire culture, researchers who have investigated its effects on team dynamics have measured it at the level of the individual. For instance, Kirkman & Shapiro (1997) found that higher collectivism on teams had a significant positive relationship with team productivity ratings from team leaders. Participants from two multi-national companies took part in the study. Researchers used Maznevski's cultural-values measure, designed specifically to measure collectivism at the level of the individual. The measure has demonstrated strong construct validity and reliability (Kirkman & Shapiro, 1997). Eight items are assessed on a 1-7 point scale of agreement, where 1=strongly disagree and 7=strongly agree. Unfortunately, this measure does not appear to be publically available. Gibson (1999) found that group efficacy was significantly related to group effectiveness for groups with high levels of collectivism.

Bell (2007) conducted a meta-analysis and found a positive correlation between collectivism and team performance, particularly when collectivism was aggregated at the team level. This is reminiscent of the personality variable research that related TPE to team performance.

While the measures of collectivism at the level of the individual used in the aforementioned studies do not appear to be available, Jackson et al., (2006) developed a measure for collectivism as an individual difference, which they call *Psychological Collectivism*. Psychological collectivism contains the following facets: Preference for in-groups, reliance on in-groups, concern for in-groups, acceptance of in-group norms, and prioritization of in-group goals. This measure has demonstrated good construct validity through its relationship with other measures of collectivism.

Instructions for this measure are as follows:

Think about the work groups to which you currently belong, and have belonged to in the past. The items below ask about your relationship with, and thoughts about, those particular groups. Respond to the following questions, as honestly as possible, using the response scales provided. (1 =Strongly Disagree to 5 =Strongly Agree).

The following are items for this measure, and their associated facet.

1. I preferred to work in those groups rather than working alone. (Preference)
2. Working in those groups was better than working alone. (Preference)
3. I wanted to work with those groups as opposed to working alone. (Preference)
4. I felt comfortable counting on group members to do their part. (Reliance)
5. I was not bothered by the need to rely on group members. (Reliance)
6. I felt comfortable trusting group members to handle their tasks. (Reliance)
7. The health of those groups was important to me. (Concern)
8. I cared about the well-being of those groups. (Concern)
9. I was concerned about the needs of those groups. (Concern)
10. I followed the norms of those groups. (Norm Acceptance)
11. I followed the procedures used by those groups. (Norm Acceptance)
12. I accepted the rules of those groups. (Norm Acceptance)
13. I cared more about the goals of those groups than my own goals. (Goal Priority)
14. I emphasized the goals of those groups more than my individual goals. (Goal Priority)
15. Group goals were more important to me than my personal goals. (Goal Priority)

### 2.4.3 *Measuring Individualism and Collectivism*

Overall, the suitability of measures for individualism and collectivism in the testbed involves an evaluation of the measure's availability, its psychometric qualities, and its applicability to the task environment and participant audience. Table 3 considers the suitability of each of these measures for inclusion in the testbed CONOPS on the basis of these factors.

Table 3. Summary of Measures of Individualism and Collectivism

| I/C Measure | Qualities | Suitability for Inclusion in CONOPS |
|---|---|---|
| *Psychological Collectivism*<br><br>Jackson, C. L., Colquitt, J. A., Wesson, M. J., & Zapata-Phelan, C. P. (2006). Psychological collectivism: A measurement validation and linkage to group member performance. *Journal of applied psychology*, *91*(4), 884. | Availability: Publically Available<br><br>Reliability and Validity: This measure has shown reasonable validity with other measures of collectivism in the literature.<br><br>Report Type: Self Report<br><br>Applicability to the Task Environment: This measure has good applicability to the task environment.<br><br>Applicability to the Target Participant: This measure refers to existing working groups, and therefore may be applicable to some participant audiences, but not to others. | **Suitable, not recommended.** This measure is readily available, is fairly compact (15 items), and has good validity. However, the items directly reference work groups the individuals already belong to. The instructions and/or items may require some adaptation depending on the audience. |
| *Measure of Cultural Values*<br><br>Maznevski, M. L., Gomez, C. B., DiStefano, J. J., Noorderhaven, N. G., & Wu, P. C. (2002). Cultural dimensions at the individual level of analysis: The cultural orientations framework. *International journal of cross cultural management*, *2*(3), 275-295. | Availability: Not publically available<br><br>Reliability and Validity: This measure was validated by Kirkman and Shapiro (1997) before the measures were published by Maznevski et al. (2002). Items showed good construct validity in that study.<br><br>Report Type: Unknown<br><br>Applicability to the Task Environment: Unknown<br><br>Applicability to the Target Participant: Unknown | **Unsuitable.** Not enough is known about this measure based on the description offered by Kirkman and Shapiro (1997). |

## 2.5 Locus of Control

Locus of Control is a construct that refers to an individual's belief regarding whether control of circumstances/events exists within a person or externally as part of the environment. "Locus" specifically references the cause of reinforcement as "internal" or "external" to the individual, and "control" references whether the cause is controllable or uncontrollable (Palenzuela, 1984). There is some reason to hypothesize that LOC is related to the constructs of individualism and collectivism discussed in the prior section, in that individualism fosters independence and reliance on the self (promoting an internal locus of control) while collectivism fosters interdependence, which necessarily distributes control of the outcome among members of a group (Spector et al., 2002). In their research regarding the constructs of individualism/collectivism, LOC, and well-being, Spector and colleagues hypothesized specifically that individualism would be associated with an internal LOC. In that work, the authors collected data on these constructs from more than 5,000 managers spread across 24 nations or territories. LOC was conceptualized as "Work Locus of Control," as measured with the Work Locus of Control Scale (WLCS) with items written either from an internal or external perspective and where high scores indicate an external LOC and lower scores indicate internal LOC (Spector, 1988). Individualism was indeed strongly correlated with an internal LOC as measured by the WLCS, and an internal LOC was positively related to measures of satisfaction and well-being (although individualism/collectivism were not related to satisfaction and well-being outright).

### 2.5.1 Operationalizing Locus of Control

LOC, while related to Individualism and Collectivism, is still a distinct and nuanced construct. In an attempt to tease out this nuance, Palenzuela emphasizes the differences between expectancy of LOC and the actual attribution of experienced successes and failures. Also as a part of this refinement of the construct, Palenzuela discusses the concept of "contingency" – the extent to which an individual feels that their results will be contingent upon their actions or attributes. Palenzuela's measure also distinguishes numerous components of non-contingency, such as luck, the influence of an external agent, an unresponsive environment, and helplessness. This particular measure does not appear to be publically available; however, a revised measure of Multidimensional Academic-Specific Locus of Control (MASLOC) was published by the same author four years later with the following dimensions: internality (contingency), helplessness (non-contingency), and luck (chance).

The directions for the MASLOC are as follows:

*In what follows you will find a series of assertions related to your PERSONAL BELIEF about certain aspects of your academic life. Your task consists of reading each assertion carefully and circling the numbers to the right of each in accordance with the score you think best fits in with the aspect in question:*

1. Completely in disagreement
2. Strongly in disagreement
3. Fairly in disagreement
4. Slightly in disagreement
5. Equally in agreement and disagreement
6. Slightly in agreement
7. Fairly in agreement
8. Strongly in agreement
9. Very much in agreement

*Please make sure that you answer all items and try to be as sincere and precise as possible in your answers, which will be kept strictly confidential.*

1. If I want to obtain a good academic record it is essential that I should have good luck.
2. The grade I get at the end of the year will always be closely related to what I do during the year.
3. Whether I get good grades or not depends on whether the random factors that may affect exams are favorable to me.

4. It is an absolute waste of time for me to make any effort since there is no relationship between my capability and how hard I work and the grades I'll get.
5. I am convinced that the grades I'll get will depend on how well or badly I do on my exams.
6. My getting good or bad grades in my exams is related to whether precisely the topics I have studied come up in the exams.
7. The kind of grades I will get in my studies depends on how capable I am of preparing myself for the subjects.
8. I don't think it is worthwhile studying hard since the grades I'll get will be completely manipulated.
9. I am convinced that whatever I do my teachers will always give me the grades *they* want to.
10. If I want to get a good academic record I have to be competent and I must work hard.
11. In general I believe that if one is competent and works hard one will get good results in one's studies.
12. Luck is something decisive in the kind of grades I get in my studies.
13. The grades I get in my subjects are always determined by a series of random circumstances.
14. It makes absolutely no difference whether I prepare well for a subject or not since in the long run the teachers are "out to catch you".
15. Regarding my academic life I just don't know what to do. Anything might happen: maybe I'll do an exam well and flunk or maybe I'll do it badly and pass.

A nuanced view of LOC was also presented by Paulhus in the 1980's, and refers to "spheres of control." This framework partitions an individual's life into three "behavioral spheres" through which the individual confronts the world. The first sphere is the non-social sphere of the "self," and has to do with personal goals the individual has. This sphere is referred to as "Personal Efficacy" (PE). The second is the sphere of "others" in dyads and group situations, and has to do with advocating for one's goals while building and maintaining social relationships. Control here refers to "Interpersonal Control" (IC). The third sphere is that of the political and social sphere, and control in this area is "Sociopolitical Control" (SC). The "Sphere of Control" (SOC) battery assessment is modeled on this framework and holds control in each of these domains as independent dispositions. In other words, an individual might have a different perception of control in one domain than another. This renders more uniform measures of LOC insufficient.

Each scale (PE, IC, SC) has ten items, worded as statements that participants rate their agreement with using a 1-7 Likert scale ranging from "disagree" to "agree."

Personal Efficacy items include:

1. When I get what I want it's usually because I worked hard for it.
2. When I make plans I am almost certain to make them work.
3. I prefer games involving some luck over games requiring pure skill.
4. I can learn almost anything if I set my mind to it.
5. My major accomplishments are entirely due to my hard work and ability.
6. I usually don't set goals because I have a hard time following through on them.
7. Competition discourages excellence.
8. Often people get ahead just by being lucky.
9. On any sort of exam or competition I like to know how well I do relative to everyone else.
10. It's pointless to keep working on something that's too difficult for me.

Interpersonal Control Scale items include:

1. Even when I'm feeling self-confident about most things, I still seem to lack the ability to control social situations.
2. I have no trouble making and keeping friends.

3. I'm not good at guiding the course of a conversation with several others.
4. I can usually establish a close personal relationship with someone I find attractive.
5. When being interviewed I can usually steer the interviewer toward the topics I want to talk about and away from those I wish to avoid.
6. If I need help in carrying off a plan of mine, it's usually difficult to get others to help.
7. If there's someone I want to meet I can usually arrange it.
8. I often find it hard to get my point of view across to others.
9. In attempting to smooth over a disagreement I usually make it worse.
10. I find it easy to play an important role in most group situations.

Sociopolitical Control Scale items include:

1. By taking an active part in political and social affairs we, the people, can control world events.
2. The average citizen can have an influence on government decisions.
3. It is difficult for people to have much control over the things politicians do in office.
4. Bad economic conditions are caused by world events that are beyond our control.
5. With enough effort we can wipe out political corruption.
6. One of the major reasons we have wars is because people don't take enough interest in politics.
7. There is nothing we, as consumers, can do to keep the cost of living from going higher.
8. When I look at it carefully I realize it is impossible to have any really important influence over what big businesses do.
9. I prefer to concentrate my energy on other things rather than on solving the world's problems.
10. In the long run we, the voters, are responsible for bad government on a national as well as a local level.

Paulhus (1983) discussed several validation studies for these subscales of spheres of control. One was a study of control profiles of varsity football players, varsity tennis players, and non-athletes. The study provided support for the construct validity of the scales because authors were able to predict the control pattern for each population based on character descriptions made by raters who knew the athletes. In fact, the PE scores appeared to be plastic and somewhat dependent on the current athletic season (tennis players had higher PE scores during tennis season when the athletes were "fully actualized" and most likely to be rewarded for their athletic ability). The interpersonal control scale may have particularly interesting implications for teaming dynamics, particularly because the IC scale was significantly relevant for athletic teams.

### 2.5.2 Measuring LOC in Teams

How does LOC factor into teaming dynamics? Lee-Kelley (2006) explored LOC as a framework for understanding how workers' general control expectancies influence attitudes about team work (particularly, distributed team work). Two hypotheses were proposed – first that those with internal and external LOC would perceive "role conflict" (the phenomenon where an individual's work deviates from the work they prefer to perform as project timelines are compressed and tasks need to be reprioritized) differently. The second was that role conflict and LOC would exhibit different relationships with job satisfaction. The researchers administered a survey to test these hypothesized relationships between constructs within a population of defense industry professionals. Participants completed the Rotter IE scale to measure LOC (an instrument comprising 23 items related to social work, political work, and educational beliefs), in addition to questionnaires on role conflict and job satisfaction. Perceptions of role conflict did not differ between those with an internal versus external LOC. However, their overall job satisfaction did, where those with a more internal LOC experienced slightly higher job satisfaction than those with an external LOC, indicating that LOC is a reasonable predictor of job satisfaction in distributed teams. Alternatively, role conflict was not, which results in the conclusion that LOC as a personality factor has more influence on an individual's satisfaction working in a distributed team than the conditions of the working environment itself. The authors also discovered an inverse relationship between job satisfaction and role conflict for individuals with an internal LOC, indicating that in the distributed team setting, those with an internal LOC tend to be affected by perceived role difficulties, which reduces their overall job satisfaction.

### 2.5.3 Measuring Locus of Control

As with other constructs, evaluating the suitability of measures of LOC in the testbed involves an evaluation of the measure's availability, psychometric qualities, and applicability to the task environment and participant audience. Table 4 considers the suitability of each of these measures for inclusion in the testbed CONOPS on the basis of these factors.

Table 4. Summary of Locus of Control Measures

| LOC Measure | Qualities | Suitability for Inclusion in CONOPS |
|---|---|---|
| **Spheres of Control**<br><br>Paulhus, D. (1983). Sphere-specific measures of perceived control. *Journal of Personality and Social Psychology*, *44*(6), 1253. | Availability: Publically Available<br><br>Reliability and Validity: Good construct validity has been demonstrated for this measure using a study of athletes (Paulhus, 1983).<br><br>Report Type: Self Report<br><br>Applicability to the Task Environment: The SOC sub-scales related to the self and interpersonal control are the most applicable to the teaming research goals anticipated for the testbed. The Sociopolitical control scale is least likely to be useful.<br><br>Applicability to the Target Participant: The items in the SOC scale should be applicable to any participant population. | **Suitable.** The individual and interpersonal subscales are most applicable for research in the testbed. |

| LOC Measure | Qualities | Suitability for Inclusion in CONOPS |
|---|---|---|
| *The Revised Multidimensional Academic-Specific Locus of Control (MASLOC)*<br><br>Palenzuela, D. L. (1988). Refining the theory and measurement of expectancy of internal vs external control of reinforcement. *Personality and Individual Differences*, *9*(3), 607-629. | Availability: Publically Available<br><br>Reliability and Validity: Palenzuela (1988) reports reasonable evidence for construct validity of the MASLOC.<br><br>Report Type: Self Report<br><br>Applicability to the Task Environment: This measure has little applicability to the task environment, because it is geared at students in a traditional educational environment.<br><br>Applicability to the Target Participant: This measure may have reasonable applicability to the participant depending on the participant population. If the participant population comprises students in a university setting, it will apply. If the participant population comprises those in occupational roles, it is not appropriate. | **Suitable, not recommended.** This measure is available, but is specifically geared toward students and not professional operators. |
| *Work Locus of Control Scale*<br><br>Spector, P. E. (1988). Development of the work locus of control scale. *Journal of occupational psychology*, *61*(4), 335-340. | Availability: Not publically available.<br><br>Reliability and Validity: The reliability and validity of this measure is unknown based on the description in Spector (1988).<br><br>Report Type: Self report<br><br>Applicability to the Task Environment: This measure appears to be related to an individual's work environment. As long as the team environment proposed by the CONOPS is similar to the individual's work environment, this should be an applicable measure of LOC.<br><br>Applicability to the Target Participant: Based on the description offered by Spector (1988) this would be a reasonable assessment to use for our target population; however it is difficult to assess because the measure is not available. | **Unsuitable.** This measure is not publically available. |

| LOC Measure | Qualities | Suitability for Inclusion in CONOPS |
|---|---|---|
| *Rotter IE Locus of Control Scale*<br><br>Lange, R. V., & Tiggemann, M. (1981). Dimensionality and reliability of the Rotter IE locus of control scale. *Journal of Personality Assessment*, *45*(4), 398-406. | Availability: Not Publically Available<br><br>Reliability and Validity: Unknown based on description available.<br><br>Report Type: Unknown based on description (presumably, this is a self-report measure).<br><br>Applicability to the Task Environment: This measure has been used widely to measure individual differences in LOC across many studies and various task environments, and therefore is probably applicable to the proposed task environment.<br><br>Applicability to the Target Participant: This measure has been used widely to measure LOC across research settings and is likely applicable to many participants. | **Unsuitable.** This measure is not publically available. |

## 2.6 Preference for Teamwork

Preference for teamwork (or preference for group work as it is sometimes labeled) is another sub-dimension of the individualism/collectivism dichotomy discussed in section 2.3. Specifically, the appeal of group work tends to be a collectivist-inclined trait, as it refers to the individual's preference for working as a part of a group as opposed to performing autonomous work (Shaw et al., 2000). It is of interest to research on teaming, as it has implications for how someone might experience satisfaction in a team environment, and, as a result, their inclination to seek out work on teams and their performance on teams.

As an example, preference for group work has been shown to interact with other aspects of group work behavior, such as individual temperament and "social loafing." Stark et al. (2007) incorporated preference for group work into a framework for understanding social loafing behavior on teams. This framework also involved constructs of competitiveness (winning orientation) and task interdependence. Stark and colleagues found that preference for group work had a negative relationship with social loafing (meaning the more people aligned with a preference to work on a team, the less likely they were to demonstrate loafing behaviors). Additionally, the desire to win in social contexts (winning orientation) moderated the negative relationship between preference for group work and social loafing, in the sense that even if they do not enjoy group work, the individual with a high winning orientation desired to work productively in a group in order to gain and maintain social advantage within the group, decoupling the relationship between a preference for that working arrangement and social loafing. Alternatively, the preference for a group work/social loafing relationship was stronger when win orientation was low.

Preference for teamwork may also affect how likely an individual is to view an autonomous agent as a tool or teammate, or how willing they are to accept information from an autonomous (compared with a human) teammate. For example, if an individual finds group work less ideal than working independently, they may be more inclined to treat an AI agent as a tool than a teammate. If the preference for teamwork is low and winning orientation is low, they may also be more likely to blindly rely on whatever information the autonomous teammate provides, instead of working collaboratively and thinking critically in order to make a meaningful contribution to the shared task. These are questions that may be interesting to explore in future human/autonomy teaming research.

### 2.6.1 Operationalizing Preference for Teamwork

Preference for team work has been primarily conceptualized as the preference to work in groups over working with other humans. As research with autonomous agents develops, this construct needs to be expanded to accommodate work with autonomous systems and agents, and geographical distribution of team members. This section will review the ways that preference for teamwork has been operationalized in human teaming studies.

The essential comparison that measures of preference for teamwork ask a respondent to make involves work in a group compared to work as an individual. Shaw et al. (2000) created a measure of preference for group work by creating a representative selection of items that reflected collectivist inclinations toward group work from a number of sources. The following seven items are factor-loaded together to support a "group work preference" factor. Agreement was recorded for each item using a seven point Likert scale.

1. When I have a choice, I try to work in a group instead of by myself.
2. I prefer to work on a team rather than individual tasks.
3. Working in a group is better than working alone.
4. Given the choice, I would rather do a job where I can work alone rather than do a job where I have to work with others in a group (R).
5. I prefer to do my own work and let others do theirs (R).
6. I like to interact with others when working on projects.
7. I personally enjoy working with others.

Shaw found that preference for group work using this measure was consistently negatively related to social loafing behavior in groups, a relationship that was moderated by a measure of winning orientation.

Several items from Shaw's (2000) measure were adapted by Luse and colleagues (2013) to measure preference for virtual teams. Those items were:

1. When I have a choice, I would rather work in virtual teams than by myself.
2. I prefer to work on a virtual team task than on individual tasks.
3. Working in a virtual group is better than working alone.
4. Given the choice, I would rather do a job where I can work alone than do a job where I have to work with others in a virtual team. (R)

While these items get at the preference for working in virtual teams over working independently, they do not address how an individual feels about virtual teams in comparison to face-to-face work. This may be of particular interest in research related to hybrid teams, since the communication style required for collaboration with autonomous agents more closely resembles the communication required by distributed teammates than co-located teammates. The researchers developed four additional items to assess this comparison:

5. I would be as comfortable working on a virtual team as I would a face-to-face team.
6. If given the appropriate technology, I can be just as effective working on a virtual team as I can on a face-to-face team.
7. I could not feel a part of a team that did not meet face-to-face. (R)

8. I would participate as easily on a team that used chat rooms, e-mail, and conference calls to communicate with my fellow team members as I could in face-to-face discussions.

Campion et al. (1993) also developed a measure of preference for team work in the context of a relatively broad analysis relating aspects of job design, interdependence, composition, context, and process to various effectiveness criteria. Campion's study accomplished three things: it 1) identified 19 work group characteristics from the relevant psychological literature that could be related to effectiveness, 2) developed a measure for those characteristics, and 3) validated that measure by relating those themes to the work of real working groups (in terms of both productivity and team member satisfaction).

Among the identified characteristics was an individual's preference for teamwork, which was operationalized with a focus on teamwork being preferable, and increasing one's ability to perform effectively. Campion et al. developed a three-item assessment with the following items:

1. If given the choice, I would prefer to work as part of a team rather than work alone.
2. I find that working as a member of a team increases my ability to perform effectively.
3. I generally prefer to work as part of a team.

Other characteristics of teamwork that are related to the construct (but not a part of the construct) measured by this study included:

- Process" (*e.g.,* "Members of my team have great confidence that the team can perform effectively")
- Social support (*e.g.,* "Being in my team gives me the opportunity to work in a team and provide support to other team members")
- Workload sharing in teams (*e.g.*, "Everyone does their fair share of the work")
- Communication and cooperation within the group (*e.g.,* "Members of my team cooperate to get the work done")

In developing their framework for understanding preference for teamwork and team-related behaviors, Stark et al. (2007) used a novel measure of preference for group work that they developed for this study. However, while the authors included one sample item from their measure, they did not provide the items in their published work, and their operationalization of the preference for group work construct is only understood as it relates to the broader framework.

### 2.6.2 *Measuring Preference for Teamwork*

Evaluating the measures of preference for teamwork for inclusion in the testbed are contingent on criteria such as the measure's availability, psychometric properties, and applicability to the research needs we anticipate encountering when using the testbed. Those considerations are summarized in Table 5.

Table 5. Summary of Preference for Teamwork Measures

| Preference for Teamwork Measure | Qualities | Suitability for Inclusion in CONOPS |
|---|---|---|
| ***Preference for Group Work***<br><br>Shaw, J. D., Duffy, M. K., & Stark, E. M. (2000). Interdependence and preference for group work: Main and congruence effects on the satisfaction and performance of group members. *Journal of Management*, *26*(2), 259-279. | Availability: Publically available.<br><br>Reliability and Validity: This measure had favorable results in a psychometric analysis reported in Shaw et al., 2000).<br><br>Report Type: Self Report<br><br>Applicability to the Task Environment: The items in this measure are worded broadly and are applicable to many task environments.<br><br>Applicability to the Target Participant: The items in this measure are worded broadly and are applicable to many participant audiences. | **Suitable.** This measure is fairly concise, has reasonable psychometric properties, and is publically available. |

| Preference for Teamwork Measure | Qualities | Suitability for Inclusion in CONOPS |
|---|---|---|
| ***Preference for Virtual Teams***<br><br>Luse, A., McElroy, J. C., Townsend, A. M., & DeMarie, S. (2013). Personality and cognitive style as predictors of preference for working in virtual teams. *Computers in Human Behavior*, *29*(4), 1825-1832. | Availability: Publically available.<br><br>Reliability and Validity: Items were adapted from Shaw et al. (2000) which has good established content validity. A confirmatory factor analysis on the adapted measure lends psychometric support for preference for virtual teams over working alone, and for virtual teams over face-to-face teaming.<br><br>Report Type: Self-report<br><br>Applicability to the Task Environment: Virtual teams are similar to distributed teams in the proposed task setting, and therefore this measure would be reasonably applicable to the task environment. However, it may be needlessly specific to the distributed aspect of teaming, rather than teaming more generally.<br><br>Applicability to the Target Participant: The items of this measure are broadly worded and the measure has applicability to many participant audiences. | **Suitable.** This measure is publically available and has good psychometric qualities. It may be too specific for a generalized measure of preference for teamwork, but would be a good addition to a testing battery. |

| Preference for Teamwork Measure | Qualities | Suitability for Inclusion in CONOPS |
|---|---|---|
| ***Preference for Teamwork***<br><br>Campion, M. A., Medsker, G. J., & Higgs, A. C. (1993). Relations between work group characteristics and effectiveness: Implications for designing effective work groups. *Personnel Psychology, 46*(4), 823-847. | Availability: Publically available.<br><br>Reliability and Validity: Campion et al., provided good support for the content validity of this measure by showing a positive relationship with employee satisfaction, but it was not tested for content validity with other similar measures.<br><br>Report Type: Self-report<br><br>Applicability to the Task Environment: The items are worded broadly and have applicability to many task environments.<br><br>Applicability to the Target Participant: The items are worded broadly and have applicability to many participant audiences. | **Suitable, not recommended.** This is a concise measure of preference for teamwork which predicted overall employee satisfaction. However, it is not validated with another form of teamwork preference or similar construct. |
| ***Preference for Teamwork***<br><br>Stark, E. M., Shaw, J. D., & Duffy, M. K. (2007). Preference for group work, winning orientation, and social loafing behavior in groups. *Group & Organization Management, 32*(6), 699-723. | Availability: Not publically available.<br><br>Reliability and Validity: Stark et al. (2007) did not elaborate in their article about the reliability or validity of their measure, but described it as being based on other previous work.<br><br>Report Type: Unknown (presumably self-report).<br><br>Applicability to the Task Environment: The measure is based on work that is thought to be widely applicable to many task environments, but this is difficult to assess without access to items.<br><br>Applicability to the Target Participant: The measure is based on work that is thought to be widely applicable to many participants, but this is difficult to assess without access to items. | **Unsuitable.** This measure is unavailable. It is also impossible to evaluate its reliability and validity based on what is presented by the authors. |

This Page Intentionally Blank

## 3 TEAM CHARACTERISTICS

The "team characteristics" family of constructs captures how well the team members relate to one another and how efficiently they work together to reach their objective. One way to think about team characteristic constructs is to categorize them on the basis of whether they are "referent shift" constructs. A referent shift construct is a single construct that exists at multiple levels – while the theory behind the construct is the same; the referent changes (Collins & Parker, 2010). Some team constructs are individual constructs that can be applied to the team (consider the individual construct of "self-efficacy," and the team construct "team efficacy"). Others are constructs that are team specific.

Another aspect of this family of constructs is that the "scores" are expected to change over time through experience. Unlike individual-level traits (the first group described in this report) that tend to be static through experience, team characteristics are dependent not only on the individuals, but also on the collective experience that a group of individuals has with one another. This is why a number of studies that examine these traits utilized multiple measurement points in their research methodology (Collins & Parker, 2010). One of the risks that Collins and Parker (2010) identify with changes over time in this kind of construct is something called "beta change," which is a "stretching" or "shrinking" of the measurement scale for items associated with these characteristics. Frequently this occurs when some kind of point of reference or standard shifts (such as graduating to new levels of expected performance, or receiving significant feedback about the team's performance). This kind of change can be detected by different factor loadings, and factor analysis can be used to investigate whether this has occurred.

### 3.1 Teamwork Attitudes and Perceptions

Teamwork attitudes and perceptions refer to the way that an individual feels about their experience working on a particular team, and their resulting attitudes about being a part of that team. Much of the work on teamwork attitudes and perceptions has been developed within the medical community, in part to develop training simulations to improve less-than-ideal teaming attitudes that may have developed over the course of one's work experience. For example, Baker et al. (2010) developed a widely used measure for teamwork attitudes in the medical field – the TeamSTEPPS Teamwork Attitude Questionnaire (T-TAQ). Motycka et al. (2018) used the T-TAQ to measure attitudes toward teamwork before and after an Event-Based Approach to Training (EBAT) exercise with the goal of improving students' attitudes regarding teamwork. The EBAT-based exercise significantly improved attitudes about one's team.

#### 3.1.1 Operationalizing Teamwork Attitudes and Perceptions

Because teamwork attitudes and perceptions are operationalized at the level (and therefore specificity) of an individual's work experience, it is difficult to find a generalized measure of this construct. This is why the measures of this construct are specific to the task environment of the target participant.

For example, the T-TAQ (described above) asks the respondent to consider their specific experience working on a particular team, rather than to consider group work more broadly. Respondents use a five-point Likert scale to indicate their agreement/disagreement with various statements[2]:

1. It is important to ask patients and their families for feedback regarding patient care.
2. Patients are a critical component of the care team.
3. This facility's administration influences the success of direct care teams.
4. A team's mission is of greater value than the goals of individual team members.

---

[2] The manual for administering the T-TAQ is available here:
https://www.ahrq.gov/teamstepps/instructor/reference/teamattitudesmanual.html

5. Effective team members can anticipate the needs of other team members.
6. High performing teams in health care share common characteristics with high performing teams in other industries.
7. It is important for leaders to share information with team members.
8. Leaders should create informal opportunities for team members to share information.
9. Effective leaders view honest mistakes as meaningful learning opportunities.
10. It is a leader's responsibility to model appropriate team behavior.
11. It is important for leaders to take time to discuss with their team members plans for each patient.
12. Team leaders should ensure that team members help each other out when necessary.
13. Individuals can be taught how to scan the environment for important situational cues.
14. Monitoring patients provides an important contribution to effective team performance.
15. Even individuals who are not part of the direct care team should be encouraged to scan for and report changes in patient status.
16. It is important to monitor the emotional and physical status of other team members.
17. It is appropriate for one team member to offer assistance to another who may be too tired or stressed to perform a task.
18. Team members who monitor their emotional and physical status on the job are more effective.
19. To be effective, team members should understand the work of their fellow team members.
20. Asking for assistance from a team member is a sign that an individual does not know how to do his/her job effectively. (R)
21. Providing assistance to team members is a sign that an individual does not have enough work to do. (R)
22. Offering to help a fellow team member with his/her individual work tasks is an effective tool for improving team performance.
23. It is appropriate to continue to assert a patient safety concern until you are certain that it has been heard.
24. Personal conflicts between team members do not affect patient safety.
25. Teams that do not communicate effectively significantly increase their risk of committing errors.
26. Poor communication is the most common cause of reported errors.
27. Adverse events may be reduced by maintaining an information exchange with patients and their families.
28. I prefer to work with team members who ask questions about information I provide.
29. It is important to have a standardized method for sharing information when handing off patients.
30. It is nearly impossible to train individuals how to be better communicators.

A related measure is the TeamSTEPPS Team Perception Questionnaire (T-TPQ), which assesses one's perception of their team. Like the T-TAQ, respondents use a five-point Likert scale to indicate their agreement with various statements[3]. These statements are grouped within several larger categories related to leadership, team structure, support, and communication, but are still worded to be very specific to the individual's current work experience.

Team Structure:

1. The skills of staff overlap sufficiently, so that work can be shared when necessary.
2. Staff are held accountable for their actions.
3. Staff within my unit share information that enables timely decision making by the direct patient care team.

---

[3] See: https://www.ahrq.gov/teamstepps/instructor/reference/teampercept.html

4. My unit makes efficient use of resources (*e.g.*, staff supplies, equipment, and information).
5. Staff understand their roles and responsibilities.
6. My unit has clearly articulated goals.
7. My unit operates at a high level of efficiency.

Leadership:

8. My supervisor/manager considers staff input when making decisions about patient care.
9. My supervisor/manager provides opportunities to discuss the unit's performance after an event.
10. My supervisor/manager takes time to meet with staff to develop a plan for patient care.
11. My supervisor/manager ensures that adequate resources (*e.g.,* staff, supplies, equipment, information) are available.
12. My supervisor/manager resolves conflicts successfully.
13. My supervisor/manager models appropriate team behavior.
14. My supervisor/manager ensures that staff are aware of any situations or changes that may affect patient care.
15. Staff effectively anticipate each other's needs.

Situation Monitoring:

16. Staff monitor each other's performance.
17. Staff exchange relevant information as it becomes available.
18. Staff continuously scan the environment for important information.
19. Staff share information regarding potential complications (*e.g.,* patient changes, bed availability).
20. Staff meets to reevaluate patient care goals when aspects of the situation have changed.
21. Staff correct each other's mistakes to ensure that procedures are followed properly.

Mutual Support:

22. Staff assist fellow staff during high workload.
23. Staff request assistance from fellow staff when they feel overwhelmed.
24. Staff caution each other about potentially dangerous situations.
25. Feedback between staff is delivered in a way that promotes positive interactions and future change.
26. Staff advocate for patients even when their opinion conflicts with that of a senior member of the unit.
27. When staff have a concern about patient safety, they challenge others until they are sure the concern has been heard.
28. Staff resolve their conflicts, even when the conflicts have become personal.

Communication:

29. Information regarding patient care is explained to patients and their families in lay terms.
30. Staff relay relevant information in a timely manner.
31. When communicating with patients, staff allow enough time for questions.
32. Staff use common terminology when communicating with each other.
33. Staff verbally verify information that they receive from one another.
34. Staff follow a standardized method of sharing information when handing off patients.
35. Staff seek information from all available sources.

Another measure of preference for teaming that calls on an individual's experience working on particular teams within their field is the "Attitudes Toward Health Care Team Scale." Muñoz de Morales-Romero et al. (2021) used this scale to measure attitudes toward teamwork before and after a six-hour training session within a high fidelity clinical simulation. The respondents used a six-point scale, from 1=strongly agree to 6=strongly agree, omitting a "neutral" option. The items are as follows:

1. Working in teams unnecessarily complicates things most of the time.
2. The team approach improves the quality of care to patients.
3. Team meetings foster communication among team members from different disciplines.
4. Physicians have the right to alter patient care plans developed by the team.
5. Patients receiving team care are more likely than other patients to be treated as whole persons.
6. A team's primary purpose is to assist physicians in achieving treatment goals for patients.
7. Working on a team keeps most health professionals enthusiastic and interested in their jobs.
8. Patients are less satisfied with their care when it is provided by a team.
9. Developing a patient care plan with other team members avoids errors in delivering care.
10. When developing interdisciplinary patient care plans, much time is wasted translating jargon from other disciplines.
11. Health professionals working on teams are more responsive than others to the emotional and financial needs of patients.
12. Developing an interdisciplinary patient care plan is excessively time-consuming.
13. The physician should not always have the final word in decisions made by health care teams.
14. The give and take among team members helps them to make better patient care decisions.
15. In most instances, the time required for team meetings could be better spent in other ways.
16. Hospital patients who receive team care are better prepared for discharge than other patients.
17. Physicians are natural team leaders.
18. The team approach makes the delivery of care more efficient.
19. The team approach permits health professionals to meet the needs of family caregivers as well as patients.
20. Having to report observations to the team helps team members to better understand the work of other health professionals.

While these measures are clearly insufficient for measuring existing attitudes and perceptions of teamwork in the planned testbed, they can provide an instructive example of how researchers can develop specific measures of teamwork attitudes that are applicable to the participants of a given study. For instance, items about leadership may be specific to mangers of healthcare providers working with patients in the T-TAQ, but could be adapted to reflect the operational environments of prospective participants.

Another set of measures are explore constructs that are not meant to provide an overall assessment of attitudes and perceptions of teamwork *per se*. Instead, they are similar to the subscales found in measures like the T-TPQ or address closely related concepts. As such, they may be adapted and combined to create a measurement battery appropriate for measuring teamwork attitudes and perceptions. For example, work by Campion et al. (1993) examined relationships between attitudes about particular team dynamics such as social support and communication and cooperation with team performance outcome variables.

Similarly, Waytz, Cacioppo, & Epley (2010) developed a measure of an individual's tendency to anthropomorphize, and demonstrated that it was a relatively stable trait. Further, the researchers demonstrated that higher levels of this trait increased the likelihood that participants would:

- Ascribe secondary emotions (*e.g.*, shame or optimism) to a non-human entity
- Express concern for a non-human entity's well-being
- Demonstrate increased trust in a non-human entity

Such a measure might be especially useful as a covariate when assessing the degree to which autonomous agents are trusted, viewed as teammates, etc.

### 3.1.2 Measuring Attitudes and Perceptions

Evaluating the measures of attitudes and perceptions of teamwork for inclusion in the testbed is contingent on criteria such as the measure's availability, psychometric properties, and (perhaps most critically for this construct) applicability to the research needs we anticipate encountering when using the testbed. Those considerations are summarized in Table 6.

Table 6. Summary of Team Attitude Measures

| Team Attitude Measure | Qualities | Suitability for Inclusion in CONOPS |
|---|---|---|
| ***Social Support within Team Settings***<br>Campion, M. A., Medsker, G. J., & Higgs, A. C. (1993). Relations between work group characteristics and effectiveness: Implications for designing effective work groups. *Personnel Psychology, 46*(4), 823-847. | Availability: Publically available.<br><br>Reliability and Validity: This measure was predictive of positive group outcomes, but was not validated using a similar measure of social support in team settings. It is also not a generalized measure of attitudes and perceptions, but specific to one aspect of team dynamics.<br><br>Report Type: Self-report<br><br>Applicability to the Task Environment: The items for this measure are broadly worded and applicable to many task environments.<br><br>Applicability to the Target Participant: The items for this measure are broadly worded and applicable to many participant audiences. | **Suitable (as component of larger battery).** This measure of social support would make a good component of a more thorough battery of related team attitude measures, but itself is not a direct measure of team attitudes. |

| Team Attitude Measure | Qualities | Suitability for Inclusion in CONOPS |
|---|---|---|
| *Communication and Cooperation within Groups*<br><br>Campion, M. A., Medsker, G. J., & Higgs, A. C. (1993). Relations between work group characteristics and effectiveness: Implications for designing effective work groups. *Personnel Psychology, 46*(4), 823-847. | Availability: Publically available.<br><br>Reliability and Validity: This measure was predictive of positive group outcomes, but was not validated using a similar measure of communication and cooperation in team settings. It is also not a generalized measure of attitudes and perceptions, but specific to one aspect of team dynamics.<br><br>Report Type: Self-report<br><br>Applicability to the Task Environment: The items for this measure are broadly worded and applicable to many task environments.<br><br>Applicability to the Target Participant: The items for this measure are broadly worded and applicable to many participant audiences. | **Suitable (as component of larger battery).** This measure of social support would make a good component of a more thorough battery of related team attitude measures, but itself is not a direct measure of generalized team attitudes. |

| Team Attitude Measure | Qualities | Suitability for Inclusion in CONOPS |
|---|---|---|
| *Individual Differences in Anthropomorphism Questionnaire (IDAQ)*<br><br>Waytz, A., Cacioppo, J., & Epley, N. (2010). Who sees human? The stability and importance of individual differences in anthropomorphism. Perspectives on *Psychological Science, 5*(3), 219-232. | Availability: Publically available.<br><br>Reliability and Validity: Złotowski et al. (2018) used the original version of the scale (30 items and an 11-point scale) and reported a Cronbach alpha of 0.92. Letheran, Kuhn, Lings, & Pope (2016) determined that a 15-item version of the original scale using a seven-point Likert scale performed well and, in fact, a nine-item version had a Cronbach alpha of 0.88.<br><br>Report Type: Self-report<br><br>Applicability to the Task Environment: The items for this measure are broadly worded and applicable to many task environments.<br><br>Applicability to the Target Participant: The items for this measure are broadly worded and applicable to many participant audiences. | **Suitable.** This measure would provide a good estimate of an individual's propensity to anthropomorphize. We plan to make the 15-item/7-point scale version available within the TCT. |

| Team Attitude Measure | Qualities | Suitability for Inclusion in CONOPS |
|---|---|---|
| ***Attitudes Toward Healthcare Team Scale***<br><br>Muñoz de Morales-Romero, L., Bermejo-Cantarero, A., Martínez-Arce, A., González-Pinilla, J. A., Rodriguez-Guzman, J., Baladrón-González, V., ... & Redondo-Calvo, F. J. (2021). Effectiveness of an Educational Intervention With High-Fidelity Clinical Simulation to Improve Attitudes Toward Teamwork Among Health Professionals. The Journal of Continuing Education in Nursing, 52(10), 457-467. | Availability: Publically available.<br>Reliability and Validity: While this measure has good psychometric properties, the items would need significant overhaul and the adapted measure would need to be re-validated.<br>Report Type: Self-report.<br>Applicability to the Task Environment: This measure has no applicability to the anticipated task environment.<br>Applicability to the Target Participant: This measure has little applicability to the anticipated target participants. | **Unsuitable.** The effort required to adapt and re-validate this measure to suit the current research needs will pose unnecessary effort. |
| ***T-TAQ***<br>Baker, D. P., Amodeo, A. M., Krokos, K. J., Slonim, A., & Herrera, H. (2010). Assessing teamwork attitudes in healthcare: development of the TeamSTEPPS teamwork attitudes questionnaire. *Quality and Safety in Healthcare, 19* (6), e49-e49. | Availability: Publically available.<br>Reliability and Validity: While this measure has good psychometric properties, the items would need significant overhaul and the adapted measure would need to be re-validated.<br>Report Type: Self-report.<br>Applicability to the Task Environment: This measure has no applicability to the anticipated task environment.<br>Applicability to the Target Participant: This measure has little applicability to the anticipated target participants. | **Unsuitable.** The effort required to adapt and re-validate this measure to suit the current research needs will pose unnecessary effort. |

| Team Attitude Measure | Qualities | Suitability for Inclusion in CONOPS |
|---|---|---|
| *T-TPQ*<br><br>Keebler, J. R., Dietz, A. S., Lazzara, E. H., Benishek, L. E., Almeida, S.A., Toor, P. A., King, H. B., Salas, E. (2014). Validation of a teamwork perceptions measure to increase patient safety. *BMJ Quality and Safety, 23*(9), 718-726. | Availability: Publically available.<br><br>Reliability and Validity: While this measure has good psychometric properties, the items would need significant overhaul and the adapted measure would need to be re-validated.<br><br>Report Type: Self-report.<br><br>Applicability to the Task Environment: This measure has no applicability to the anticipated task environment.<br><br>Applicability to the Target Participant: This measure has little applicability to the anticipated target participants. | **Unsuitable.** The effort required to adapt and re-validate this measure to suit the current research needs will pose unnecessary effort. |

### 3.2 Team Efficacy

The concept of team efficacy can be described as a team member's perceived confidence in their team's abilities. In research, it has been conceptualized by a few constructs. For example, the construct of "team potency" has been described as the general ability of a team to achieve tasks (Collins & Parker, 2010). Prior research has demonstrated a positive relationship between team potency and performance, although the construct is believed to relate more with a team member's positive affect than their cognitive ability to perform work (Collins & Parker, 2010). A way of representing perceptions of team efficacy more precisely is by considering the processes teams use to achieve their goal, and the outcomes their team experiences. For example, team efficacy research conducted by Collins and Parker (2010) distinguishes between processes and outcomes by defining "team outcome efficacy" as the belief that the team can successfully execute a particular set of tasks, and "team process efficacy" as beliefs that team members have about their ability to work collectively. Related to this outcome/process distinction is Hu and Linden's (2011) measure of "goal clarity" and "process clarity." Goal clarity refers to the clarity with which a team member understood their objectives on the team. Process clarity is the confidence that someone has in their ability to schedule activities in their day. Yet another construct developed to understand a team's perception of their performance ability is by measuring "collective efficacy," which is the belief that an individual holds about their group's ability to successfully perform their work. "Collective outcome expectancy" refers to the beliefs an individual holds about the likely consequences the group will experience as a result of the group's performance of work tasks (Riggs & Knight, 1994). This section will review the operationalization of team efficacy through process and outcome oriented measures, and review the suitability of those measures.

### *3.2.1 Operationalizing Team Efficacy*

Collins and Parker (2010) have conducted significant research on the construct of "team potency" as a way of assessing an individual's belief in their team's ability. Their work hypothesized that team potency creates a stimulating and rewarding work environment for team members, thereby increasing the creative and cognitive efforts exerted by team members (Collins & Parker, 2010). Collins and Parker (2010) also predicted that notions of team potency would change over time, based on an individual's experience within a team. Their study involved two samples of executives in a MBA program working in teams of 4-6 team members. The researchers measured team potency using a measure created by Guzzo et al., (1993). This measure had respondents rate the following statements using a 5-point Likert (1="to no extent" to 5= "to great extent"):

- This team believes it can become unusually good at producing high quality work.
- This team expects to be known as a high performing team.
- My team feels like it can solve any problem it encounters.
- My team has confidence in itself.
- My team believes it will get a lot done when it works hard.
- No task is too tough for this team.
- My team believes it can be very productive.
- My team expects to have a lot of influence in the classroom.

Team performance was measured as a score out of 100 for an assignment completed by teams at the 6-week mark. Collins and Parker (2010) found that this measure was less predictive than more precise measures of team efficacy, but advocated for the potential value of the team potency construct in cases where the team's task does not have a clear metric for success, or concrete performance feedback is unrealistic.

Similar to team potency (and used as a proxy for team potency; see Hu & Liden, 2011) Riggs and Knight (1994) created a measure for a construct they called "collective efficacy." The collective efficacy measure involved a response to the following items using a six-point scale (1 = strongly disagree to 6 = strongly agree):

1. The department I work with has above average ability.
2. This department is poor compared to other departments doing similar work. (R)
3. This department is not able to perform as well as it should. (R)
4. The members of this department have excellent job skills.
5. Some members of this department should be fired due to lack of ability. (R)
6. This department is not very effective. (R)
7. Some members in this department cannot do their jobs well. (R)

While these sole measures of efficacy and potency have been useful in the literature, describing team efficacy using multiple constructs appears to be a more precise way of measuring it. For example, Collins and Parker (2010) found that measures of outcome and process efficacy were more effective than team potency in predicting concrete measures of team success. In research on team efficacy that delineates between outcome and process measures, these constructs tend to be operationalized as an individual's confidence in their team's ability to work together effectively to achieve shared goals. Collins and Parker (2010) created measures for team outcome efficacy and team process efficacy using an 11-point Likert scale ranging from "unconfident" to "totally confident" in their measures. The development of the measure for team outcome efficacy was developed by adapting a version of Bandura's (2006) traditional "self-efficacy" scale, with items modified to refer to a team task (*i.e.*, the confidence of earning a particular grade on a group project). The item stem was formulated as "In your current team assignment, how confident are you that your team can achieve a grade of at least ___". The stem was completed by 11 items of increasing score – 50%, 50-54%, and so on in increments of 5% until reaching 100%.

Measuring team process efficacy required the authors to create a new measure to assess an individual's confidence of the team's motivation at the present time. The researchers developed a 10-item measure of team process efficacy. Items associated with team process efficacy were one's confidence that they would:

- Approach members who are overexerting or misusing their influence.
- Resolve conflicts that have become personalized.
- Confront members who behave in unacceptable ways.
- Identify realistic goals that unify individual team member goals.
- Listen with an open mind to team member ideas.
- Seek feedback on how well your team is performing.
- Actively help team members.
- Adapt to changing situations/demands.
- Coordinate team members so they contribute their unique skills and abilities.
- Build positive morale among team members.

Collins and Parker (2010) found that team outcome efficacy was the strongest predictor of team performance in their study, while team process efficacy was shown to have a stronger relationship with the construct of "team citizenship" (the notion that members of the team are creating a teaming environment conducive to team effectiveness).

An emphasis on team processes is also explored in Hu and Liden's research (2011). Hu and Liden examined three concepts: process clarity, goal clarity, and servant leadership. They investigated how these concepts contributed to perceptions of team potency. Goal and process clarity were each assessed using a five-item scale developed by Sawyer (1992), where respondents were asked to indicate how certain they were about each aspect of their work (from 1=very uncertain, to 6=very certain).

The items associated with process clarity were to rate one's certainty of the following:

- How to divide my time among the tasks required of my job.
- How to schedule my work day.
- How to determine the appropriate procedures for each work task.
- The procedures I use to do my job are correct and proper.
- Considering all your work tasks, how certain are you that you know the best ways to do these tasks?

Unlike the construct of team process efficacy described by Collins and Parker (2010), Hu and Liden's (2011) measure of process clarity is more related to an individual's confidence in how they need to perform independently on their work, rather than how comfortable they are interacting productively with their teammates. Both are important for successful team performance, and, in fact, Hu and Liden (2011) found that process clarity was positively related to Riggs and Knight's (1994) measure of collective efficacy.

As described in several sections earlier in this report, Campion et al (1993) also developed a number of self-report measures that assessed a team member's attitudes about the ability of their team to perform on various dimensions. One of these dimensions was team potency, with the following items:

1. Members of my team have great confidence that the team can perform effectively.
2. My team can take on nearly any task and complete it.
3. My team has a lot of team spirit.

However, other constructs measured by Campion et al. (1993) also capture relevant team dynamics that may be useful in measuring team potency, such as the team's ability to manage itself, participate in team decisions, and share the workload.

The items for this scale were as follows:

Self-Management:

1. The members of my team are responsible for determining the methods, procedures, and schedules with which the work is done.
2. My team rather than my manager decides who does what tasks within the team.
3. Most work-related decisions are made by the members of my team rather than by my manager

Participation:

1. As a member of a team, I have a real say in how the team carries out its work.
2. Most members of my team get a chance to participate in decision making.
3. My team is designed to let everyone participate in decision making.

Workload Sharing:

1. Everyone on my team does their fair share of the work.
2. No one in my team depends on other team members to do the work for them.
3. Nearly all the members on my team contribute equally to the work.

In Campion's study, self-management and participation constructs all showed positive relationships with most of the effectiveness criteria and were the most predictive of effectiveness. Team potency, workload sharing, and communication/cooperation largely predicted productivity and judgments of managers. Team potency itself was a strong predictor of all three measurement criteria.

### 3.2.2 *Measuring Team Efficacy*

Evaluating the measures of team potency for inclusion in the testbed is contingent on criteria such as the measure's availability, psychometric properties, and applicability to the research needs we anticipate encountering when using the testbed. Those considerations are summarized in Table 7.

Table 7. Summary of Team Efficacy Measures

| Team Efficacy Measure | Qualities | Suitability for Inclusion in CONOPS |
|---|---|---|
| *Self-Management*<br><br>Campion, M. A., Medsker, G. J., & Higgs, A. C. (1993). Relations between work group characteristics and effectiveness: Implications for designing effective work groups. *Personnel Psychology, 46*(4), 823-847. | Availability: Publically available.<br><br>Reliability and Validity: This measure was predictive of positive group outcomes, but was not validated using a similar measure of team self-management. It is also not a generalized measure of team potency, but specific to one aspect of team dynamics that may be related to potency.<br><br>Report Type: Self-report<br><br>Applicability to the Task Environment: The items for this measure are broadly worded and applicable to many task environments.<br><br>Applicability to the Target Participant: The items for this measure are broadly worded and applicable to many participant audiences. | **Suitable (as component of larger battery).** This measure of self-management would make a good component of a more thorough battery of related team attitude measures, but itself is not a direct measure of team potency. |

| Team Efficacy Measure | Qualities | Suitability for Inclusion in CONOPS |
|---|---|---|
| *Workload Sharing*<br><br>Campion, M. A., Medsker, G. J., & Higgs, A. C. (1993). Relations between work group characteristics and effectiveness: Implications for designing effective work groups. *Personnel Psychology, 46*(4), 823-847. | Availability: Publically available.<br><br>Reliability and Validity: This measure was predictive of positive group outcomes, but was not validated using any similar measure of work load. It is also not a generalized measure of team potency, but specific to one aspect of team dynamics that may be related to potency.<br><br>Report Type: Self-report<br><br>Applicability to the Task Environment: The items for this measure are broadly worded and applicable to many task environments.<br><br>Applicability to the Target Participant: The items for this measure are broadly worded and applicable to many participant audiences. | **Suitable (as component of larger battery).** This measure of work load would make a good component of a more thorough battery of related team attitude measures, but itself is not a direct measure of team potency. |

| Team Efficacy Measure | Qualities | Suitability for Inclusion in CONOPS |
|---|---|---|
| *Participation*<br><br>Campion, M. A., Medsker, G. J., & Higgs, A. C. (1993). Relations between work group characteristics and effectiveness: Implications for designing effective work groups. *Personnel Psychology, 46*(4), 823-847. | Availability: Publically available.<br><br>Reliability and Validity: This measure was predictive of positive group outcomes, but was not validated using any similar measure of inter-team participation. It is also not a generalized measure of team potency, but specific to one aspect of team dynamics that may be related to potency.<br><br>Report Type: Self-report<br><br>Applicability to the Task Environment: The items for this measure are broadly worded and applicable to many task environments.<br><br>Applicability to the Target Participant: The items for this measure are broadly worded and applicable to many participant audiences. | **Suitable (as component of larger battery).** This measure of inter-team participation would make a good component of a more thorough battery of related team potency measures, but itself is not a direct measure of team potency. |

| Team Efficacy Measure | Qualities | Suitability for Inclusion in CONOPS |
|---|---|---|
| *Team Potency Scale*<br><br>Guzzo, R. A., Yost, P. R., Campbell, R. J., & Shea, G. P. (1993). Potency in groups: Articulating a construct. British Journal of Social Psychology, 32, 87–106. | Availability: Items publically available (development publication unavailable).<br><br>Reliability and Validity: Prior studies have offered evidence for good reliability for this measure. The content validity of this measure is unknown because the original article reporting on the development of the measure was unavailable.<br><br>Report Type: Self-report<br><br>Applicability to the Task Environment: The items are worded broadly and have wide applicability to many task environments.<br><br>Applicability to the Target Participant: The items are worded broadly and have wide applicability to many participant audiences. | **Suitable, not recommended.** This measure has acceptable psychometric properties, with the exception of the unknown level of content validity. More needs to be understood about validity before this measure can be recommended for use. |
| *Team Potency*<br><br>Campion, M. A., Medsker, G. J., & Higgs, A. C. (1993). Relations between work group characteristics and effectiveness: Implications for designing effective work groups. *Personnel Psychology, 46*(4), 823-847. | Availability: Publically available.<br><br>Reliability and Validity: This measure was predictive of positive group outcomes, but was not validated using a similar measure of team potency.<br><br>Report Type: Self-report<br><br>Applicability to the Task Environment: The items for this measure are broadly worded and applicable to many task environments.<br><br>Applicability to the Target Participant: The items for this measure are broadly worded and applicable to many participant audiences. | **Suitable, not recommended.** This measure has not been validated with any other similar measure of team potency. However, it was predictive of overall group performance. |

| Team Efficacy Measure | Qualities | Suitability for Inclusion in CONOPS |
|---|---|---|
| *Collective Efficacy*<br><br>Riggs, M. L., & Knight, P. A. (1994). The impact of perceived group success-failure on motivational beliefs and attitudes: A causal model. *Journal of Applied psychology*, *79*(5), 755. | Availability: Publically available.<br><br>Reliability and Validity: The collective efficacy scale demonstrated good reliability in Riggs and Knight's (1994) study. The scales demonstrated good reliability and predictive validity with measures performance. Factor analysis demonstrated collective efficacy as a distinct construct.<br><br>Report Type: Self-report<br><br>Applicability to the Task Environment: The items in this measure are worded to address beliefs about a corporate environment, not a mission-oriented team in a high stakes environment, and therefore has limited applicability to the task environment.<br><br>Applicability to the Target Participant: The items in this measure are worded to address beliefs about a corporate work environment. Since we do not anticipate participants with familiarity in that environment, this measure has limited applicability to target participant groups. | **Suitable, not recommended.** This measure is worded to apply to a corporate environment, which limits its utility for inclusion in the testbed. |

## 3.3 Team Cohesion

Team cohesion has proven to be a challenging construct to study over decades of research dedicated to understanding it and its relationship to team efficacy. This problem is at least partially due to little agreement that researchers have on what the construct is; an operational mechanism and a theory backing that mechanism are largely missing (Casey-Campbell & Martins, 2009). In general, it is regarded as how inclined group members are to forge social bonds, resulting in members sticking together and remaining united. However, it is also difficult to separate the definition of group cohesion from the antecedents of group cohesion (the things that cause it). For instance, Festinger's (1950) foundational definition of cohesiveness describes it as "…the resultant of all the forces acting on members to remain in the group." In considering this definition, Casey-Campbell and Martins (2009) point out that it could include individual traits possessed by team members before the group formed (*i.e.*, the trait-level factors we described in section 2), in combination with factors that developed as a consequence of social interaction within the group.

### 3.3.1 Operationalizing Team Cohesion

From a measurement standpoint, operationalizing team cohesion is a challenge in the teaming literature because the way it tends to be operationalized is specific to the domain in which it is measured (Casey-Campbell & Martins, 2009). Therefore, measures of cohesion do not have universal application to teams in general. However, two general components of cohesion that are assumed to be related to team performance are improved communication among team members and increased participation among team members.

From a methodological standpoint, there are two methods for operationalizing team cohesion in the research literature. The first is to define it as a trait of teams, examine distinct groups with low and high levels of cohesion, and look for performance differences between the groups (this can be done by deliberately creating those groups within the experiment). The other is to consider team cohesion as a continuous variable, and assess the correlation between cohesion and performance (Casey-Campbell & Martins, 2009). The trouble with this, however, is that the direction of the relationship between cohesion and performance cannot be well-understood using this paradigm. For example, a study by Stashevsky and Koslowsky (2006) looked at the interactive effects of leadership, gender, knowledge level, and team cohesion on team performance. In this study, the researchers developed a fairly generic novel three-item scale to assess team cohesion, where respondents used a five point Likert type scale to report their agreement with each sentence:

1. The contributions of the team members were equal.
2. The team atmosphere was good.
3. The team decisions were participative.

Both team member knowledge and team cohesion positively predicted overall team performance. However, the direction of the relationship between cohesion and performance is not clear in the correlational method of assessing this relationship. One is not able to conclude from this finding that the correlation is produced by a positive effect of cohesion on performance. In fact, there is some evidence that successful team performance improves cohesion (Casey-Campbell & Martins, 2009).

One important aspect up for debate in the operationalization of team cohesion is the specific dimensions implicated in the construct. A popular way of conceptualizing cohesion is by considering both "social" and "task" dimensions (Caron et al., 2003). Various researchers have found different ways that these construct dimensions impact teaming. For example, Zaccarow and Lowe (1988) examined the relationships among task-based and interpersonal cohesiveness and group performance. Task-based cohesion refers to the sense that group membership helps an individual attain personal goals, while interpersonal cohesion is the extent to which members of the group have positive relationships with one another. In this study, individual members made products called "moon tents." The group's performance was a summation of all individual products created by team members (something referred to as "pooled workflow" in the teaming literature; see Saavedra, Earley & Van Dyne, 1993). Performance on this type of task is maximized when individuals exert maximum effort on their part of the task, and members of the team have minimal interactions that might interfere with their individual work. Within this type of task, the researchers hypothesized that task-based cohesion would improve team performance, but that social cohesion may diminish performance (the alternative being that if cohesion does not have separate task and social dimensions, then cohesion overall would support team performance). The researchers manipulated social cohesion by exposing teams to training that either enhanced their relationships or seeded a sense of distance among team members, and manipulated task-related cohesion by offering extra credit to teams who produced the most moon tents. To conduct a manipulation check, the researchers developed two measures. Task cohesion was measured by asking participants to indicate:

1. The degree to which they felt success was personally important.
2. The amount of effort they believed they would expend completing a task.
3. The degree to which doing well in the experiment was more or less important than doing well in the average psychology experiment in which they had participated.
4. The degree to which subjects would recommend participation in the experiment to a friend.
5. The perceived applicability of the experiment's results.
6. The perceived personal benefit accrued by participating in the experiment.

Interpersonal cohesion was assessed by asking group members to rate their group on the following using 11-point scales:

- Cold-warm
- Unpleasant-pleasant
- Dislikable-likable
- Courteous-discourteous
- Undependable-dependable
- Friendly-unfriendly
- Bold-curious
- Casual-deliberate
- Liberal-conservative
- Nonchalant-serious
- Whether they felt their group was "close"

Zaccaro and Lowe's (1988) results revealed that manipulating task and social cohesion had different effects on teammates' perceptions of team cohesion. For example, those in the high task cohesion/low social cohesion group reported liking their teammates significantly less than those who were in the low task cohesion/low social cohesion group. In other words, engendering a sense of task cohesion appeared to result in less social cohesion than would have been present without the task cohesion intervention. Overall, participants in the high task cohesion group produced more moon tents than groups with low task cohesion. Interpersonal cohesion appeared not to affect team performance at all. However, there are some issues with Zaccaro and Lowe's method. For one, social cohesion was manipulated to be "high" or "low" without an adequate control condition. Thus, one could argue they compared high social cohesion with a sense of dissimilarity rather than a sense of neutrality toward teammates. This is problematic for interpreting the task by the social cohesion interaction mentioned above. Second, the task required very little team interdependence, which poses a significant issue for the applicability of the task environment across many team settings. Social cohesion might have had an impact on team performance if the task required more interdependence, such as a reciprocal model requiring "give and take" among adjacent team members, or a team model that relies on information, resource, and work product sharing among everyone on the team.

Another way dimensions of cohesion have been conceptualized is to define constructs of "group integration," which refers to a group member's perception of the group overall, and "individual attraction to the group" referring to how attracted that individual is to being a part of that group (Caron et al., 1985). In fact, Caron et al. (1985) devised their model of team cohesion by starting with the group integration and attraction dimensions, and sub-dividing those into social and task-based constructs. Chang and Bordia (2001) conducted a study to assess Caron's four-part factor structure of cohesion at a group level.

Participants were students in an organizational psychology course completing a five-week group project. The time-one measure was taken in the second week of the project, and the time-two measure was taken during the fifth week. Researchers used a modified set of the group integration items from the Group Environment Questionnaire (GEQ) (Caron et al., 1985). Since the GEQ was designed for athletic teams initially, the wording of the items was changed from references to sports teams to project-related work, working hours, and project outcomes. Task cohesion was measured with four items from the GEQ, and social cohesion was measured by three GEQ items. Group effectiveness was evaluated using the following measures:

- A group presentation grade (out of 10).
- A five point scale of productivity (1=not productive to 5=very productive).
- A subjective measure of how well they worked as a group (1=very poor to 5=very good).
- System viability (referring to how much they enjoyed the work; 1=very little to 5=very much).
- Self-reports of professional growth (one indicating how much technical knowledge they learned; 1=not much at all to 5=very much.
- How much the group project helped them understand how to work as a team (1=not much at all to 5=very much).

Chang and Bordia used Principal Component Analysis (PCA) to examine the structure of cohesion and group performance, and regression analyses were used to examine the relationships between these constructs. The PCA on cohesion provided evidence for a two-factor solution: task and social cohesion. The PCA for performance offered some initial support for a three-factor structure (mapping to the self-report measures of performance mentioned earlier).

In the analysis of relationships between cohesion and performance, the researchers conceptualized both constructs at the group level, so the researchers aggregated the data at the group level. In the regression analyses conducted at time one, task cohesion significantly predicted subjective measures of group performance and professional growth, while social cohesion significantly predicted system viability. At the second measurement time, task cohesion was the only significant predictor for a subjective measure of group performance. Task and social cohesion predicted system viability, but neither predicted professional growth. Overall, task cohesion was a strong predictor for a subjective measure of group cohesion, and social cohesion positively predicted system viability. Notably, task cohesion was not a strong predictor of group grade (the only non-subjective performance measure). Additionally, this study did not find any support for the idea that subjective measures of team performance positively influenced either types of cohesion.

Carron et al. (1985) developed their model of cohesion measured by the GEQ. These factors included 1) the individual group members' perceptions of the group's wholeness and 2) the individual members' personal concerns about remaining within the group, both of which could be further divided into "social" and "task-related" dimensions. Carless and De Paola (2000), however, found support for a three-factor model using the GEQ, which involved the following dimensions: task cohesion, social cohesion, and attraction to the group. The results of their work modified the GEQ from 18 items to 10 items (see Table 8).

**Table 8: Ten-item GEQ from Carless and De Paola (2000)**

| Item | Reverse Scored? | Sub-scale |
|---|---|---|
| Our team is united in trying to reach its goals for performance | No | Task Cohesion |

| Item | Reverse Scored? | Sub-scale |
|---|---|---|
| I'm unhappy with my team's level of commitment to the task | Yes | Task Cohesion |
| Our team members have conflicting aspirations for the team's performance | Yes | Task Cohesion |
| This team does not give me enough opportunities improve my personal performance | Yes | Task Cohesion |
| Our team would like to spend time together outside of work hours | No | Social Cohesion |
| Members of our team do not stick together outside of work time | Yes | Social Cohesion |
| Our team members rarely party together | Yes | Social Cohesion |
| Members of our team would rather go out on their own than get together as a team | Yes | Social Cohesion |
| For me this team is one of the most important social groups to which I belong | No | Individual Attraction to the Group |
| Some of my best friends are in this team | No | Individual Attraction to the Group |

Another controversy surrounding the operationalization of team cohesion is establishing the appropriate level of measurement. In their review article on group cohesion, Casey-Campbell and Martins (2009) reviewed several ways that cohesion can be measured:

- At the level of the individual
- At the level of the group (*e.g.,* Carless & DePaola, 2000)
- Both (*e.g.,* Carron et al., 1985)
- As aggregated individual perceptions

Measuring cohesion at the level of the individual is straightforward. Measuring at the level of the group is somewhat more complicated because there are fewer methods that can be used. One is to aggregate beliefs of individuals. Another is to reach consensus on survey items, which requires group members to work together (meaning they can influence one another). A third method is to use individuals as informants on collective group beliefs (as Caron et al. (2003) do in their study). There is some question as to whether individual self-reported beliefs can accurately reflect the beliefs of the group. Caron et al. (2003) describe a statistical method known as the "index of agreement," which is an index of the extent to which group members are in agreement about a particular concept. The researchers calculated this index on a collection of athletic teams' responses to the GEQ. The 18-item GEQ had been collected for all team members, which assessed four dimensions of cohesion: task oriented individual attraction to the group, social oriented individual attraction to the group, task oriented group integration, and social oriented group integration. The researchers calculated the non-adjusted index of agreement, which is the systematic variance in judgments about a target stimulus, and the adjusted index of agreement, which takes into account a biased distribution. Agreement on the four types of cohesion varied widely among teams. The highest degree of shared beliefs tended to be on measures of group integration (for both task and social dimensions). Individual attraction to the group for the social dimension had the least amount of agreement among team members, and overall both task and social dimensions of individual attraction to the group had the lowest inter-group consensus. Caron et al.'s study is an interesting illustration of how different types of cohesion can manifest among teams while offering a method to calculate cohesion among a group using a measure at the individual level. This is also a potentially interesting method of statistical validation of cohesion measures, as agreement index could be compared among multiple measures of group cohesion.

Although distinct from cohesion, the degree to which one individual views another as a "teammate" is a closely aligned concept. Further, this construct has significant meaning in Human-AI teaming settings. Wynne and Lyons (2018; 2019) developed the Autonomous Agent Teammate-likeness (AAT) measure for just this context. The AAT uses five-point Likert scale items to assess a given entity along six dimensions:

1. Perceived agentic capability
2. Perceived benevolence/altruistic intent
3. Perceived task interdependence
4. Perceived relationship-oriented behavior
5. Perceived richness of communication
6. Perceived presence of a mind and a shared mental model

It may be useful to revisit the AAT measure in a research setting that adjusts the features of an AI entity. It would be interesting to investigate the degree to which designers can use certain techniques (*e.g.*, adopting an animal form factor *vs*. a human form factor, controlling the apparent politeness of the automated system) to moderate human expectations and behaviors. Could techniques be used to reduce the ascription of a mind and intentionality to an artificial entity, and thereby prevent surprises, loss of trust, and disuse (Kapalo, Phillips, & Fiore, 2016; Phillips, Schaefer, Billings, Jentsch, & Hancock, 2016; McNeese, Flathmann, O'Neill, & Salas, 2023)? Would that lead to superior system performance or a generalized lack of use? This work could be used to refine the AAT measure and explore its interaction with the anthropomorphizing trait measure developed by Waytz, Cacioppo, & Epley (2010).

### 3.3.2 *Measuring Team Cohesion*

In summary, we identified five measures of team cohesion in our review of the literature, which can be found in Table 9.

Table 9. Summary of Team Cohesion Measures

| Team Cohesion Measure | Qualities | Suitability for Inclusion in CONOPS |
|---|---|---|
| *10-item Revised GEQ*<br>Carless, S. A., & De Paola, C. (2000). The measurement of cohesion in work teams. *Small Group Research, 31*(1), 71-88. | Availability: Publically available.<br><br>Reliability and Validity: Carless and DePaola's (2000) work validated their revised measure in their study (reliability was not assessed).<br><br>Report Type: Self-report<br><br>Applicability to the Task Environment: The items of this measure are reasonably broad and could support a lot of different teaming environments.<br><br>Applicability to the Target Participant: One drawback is that the questions on this measure are specific to teams that are co-located, and so would not apply to distributed teams (which may be important for the research conducted in the proposed testbed). | **Suitable.** This measure modified and validated the original GEQ, making it better suited to teams of all kinds, and solidifying a multi-factor model of team cohesion. |

| Team Cohesion Measure | Qualities | Suitability for Inclusion in CONOPS |
|---|---|---|
| ***Autonomous Agent Teammate-likeness (AAT)***<br><br>Wynne, K. T., & Lyons, J. B. (2018). An integrative model of autonomous agent teammate-likeness. Theoretical Issues in Ergonomics Science, 19(3), 353-374.<br><br>Wynne, K. T., & Lyons, J. B. (2019). Autonomous Agent teammate-likeness: scale development and validation. In Virtual, Augmented and Mixed Reality. Applications and Case Studies: 11th International Conference, VAMR 2019, Held as Part of the 21st HCI International Conference, HCII 2019, Orlando, FL, USA, July 26–31, 2019, Proceedings, Part II 21 (pp. 199-213). Springer International Publishing. | Availability: Publically available.<br><br>Reliability and Validity: Each of the six ATT subscales has adequate reliability (0.75 – 0.91). The dimensions were positively correlated, but multi-collinearity was absent, indicating that the dimensions captured distinct constructs<br><br>Report Type: Self-report<br><br>Applicability to the Task Environment: The items of this measure are reasonably broad and could support a lot of different teaming environments.<br><br>Applicability to the Target Participant: The measure is broadly applicable to a range of teaming environments. | **Suitable.** This measure should provide a very useful measure of how teammate-like human participants feel various agents are. |

| Team Cohesion Measure | Qualities | Suitability for Inclusion in CONOPS |
|---|---|---|
| **Group Environment Questionnaire (GEQ)**<br><br>Carron, A. V., Widmeyer, W. N., & Brawley, L. R. (1985). The development of an instrument to assess cohesion in sport teams: The Group Environment Questionnaire. *Journal of Sport and Exercise Psychology* | Availability: Not publically available.<br><br>Reliability and Validity: The content and construct validity of this measure was supported in Carron and Widmeyer's (1985) study. However, Casey-Campbell and Martins (2009) note the measure's questionable reliability and validity.<br><br>Report Type: Self report<br><br>Applicability to the Task Environment: This measure is specifically designed for use in athletic teams, and therefore would need to be revised to fit the relevant task environment.<br><br>Applicability to the Target Participant: This measure is not applicable to the target participants for the testbed, and would need to be heavily revised to apply to target participants. | **Suitable, not recommended.** While this has been a popular measure of team cohesion in the literature, it is tailored specifically to application in athletic domains. Furthermore, Casey-Campbell and Martins (2009) note the measure's questionable reliability and validity. |

| Team Cohesion Measure | Qualities | Suitability for Inclusion in CONOPS |
|---|---|---|
| ***Revised Task and Social Cohesion Subscales from GEQ***<br>Chang, A., & Bordia, P. (2001). A multidimensional approach to the group cohesion-group performance relationship. *Small Group Research, 32*(4), 379-405. | Availability: Item stems are publically available (would need to contact authors for specific wording).<br><br>Reliability and Validity: Validity of these subscales is demonstrated through its positive predictive relationship to team performance. The modified measure aligned with the same dimensions of the original GEQ, and had good internal consistency.<br><br>Report Type: Self-report<br><br>Applicability to the Task Environment: The items were modified to reflect the specific task used by this study, and are not applicable to the testbed.<br><br>Applicability to the Target Participant: The items could apply broadly to many participants, but were worded to be task-specific. | **Suitable, not recommended.** The modified subscales from task and social cohesion positively predicted subjective measures of team performance, but were specifically tailored to the nature of the work the group in the study was engaged in. These subscales will be more valuable for our research purposes if modified to fit the nature of the task environment. |

| Team Cohesion Measure | Qualities | Suitability for Inclusion in CONOPS |
|---|---|---|
| **3-item Novel measure of Cohesion**<br>Stashevsky, S., & Koslowsky, M. (2006). Leadership team cohesiveness and team performance. *International Journal of Manpower*. | Availability: The items for this measure are publically available.<br>Reliability and Validity: This measure was positively correlated with team performance. A reliability analysis was not reported by the study. Importantly, this measure does not capture the multi-dimensional nature of the construct.<br>Report Type: Self-report<br>Applicability to the Task Environment: The items are broadly applicable to a range of teaming scenarios, including the task environment of the testbed.<br>Applicability to the Target Participant: The items are broadly applicable to a range of participants, including the target participants for the testbed. | **Unsuitable.** Our review of the literature in team cohesion revealed a multi-dimensional model for team cohesion that was not supported by the novel measure devised for this study. |

| Team Cohesion Measure | Qualities | Suitability for Inclusion in CONOPS |
|---|---|---|
| ***Task-based Cohesion and Social Cohesion Manipulation Check Questionnaires***<br><br>Zaccaro, S. J., & Lowe, C. A., (1988). Cohesiveness and performance on an additive task: Evidence for the multidimensionality. *The Journal of Social Psychology, 128*(4), 547-558. | Availability: Questionnaires are publically available.<br><br>Reliability and Validity: These questionnaires were developed to be used as a manipulation check, and reliability and validity analyses were not conducted.<br><br>Report Type: Self-report<br><br>Applicability to the Task Environment: Task-based cohesion items were specific to the nature of this task being a psychological experiment and would not apply to the testbed. The items for the social cohesion measure are more broadly applicable to a range of task environments.<br><br>Applicability to the Target Participant: The items for both questionnaires are broadly applicable to many participants, including the target participants for the testbed. | **Unsuitable.** These questionnaires were designed to confirm the validity of an experimental manipulation on both types of cohesion, and do not have the psychometric validity of other available measures. |

# 4 OVERALL SUMMARY OF RECOMMENDED VARIABLES AND MEASURES

Two categories of constructs were reviewed for this report as having shown having significant influence on team performance. The first category reviewed was that of individual trait constructs, which focuses on stable characteristics that make up an individual's temperament. Measures in this category included measures of personality, predisposition to individualism or collectivism, an individual's interest in working on teams, and LOC. The recommended measures of consideration in the CONOPS are summarized in Table 10.

Table 10. Summary of Individual Trait Constructs to be Included in TCT CONOPS

| Recommended Measure | Measurement Qualities |
|---|---|
| **Big Five Personality Traits** | |
| IPIP-NEO-120 (Johnson, 2014) | Availability: Publically available |
| | Reliability and Validity: Validated and used across multiple psychological areas of research (see Johnson, 2014). |
| | Report Type: Self Report |
| | Applicability to the Task Environment: This measure is fairly universally applicable and reasonable for the task environment. |
| | Applicability to the Target Participant: This measure is fairly universally applicable and reasonable for the audience. However, it is a longer measure involving 120 items and may be cumbersome to involve in research. |
| Mini IPIP (Donnellan, Oswald, Baird & Lucas, 2006) | Availability: Publically Available |
| | Reliability and Validity: Validated and used across multiple psychological areas of research. |
| | Report Type: Self report |
| | Applicability to the Task Environment: This measure is fairly universally applicable and reasonable for the task environment. |
| | Applicability to the Target Participant: This measure is fairly universally applicable and reasonable for the audience. |

| Recommended Measure | Measurement Qualities |
|---|---|
| **Emotional Intelligence** | |
| Trait Emotional Intelligence Questionnaire (TEIQue) (Petrides, 2009) | Availability: Publically Available<br><br>Operationalization Variant: Trait measure<br><br>Reliability and Validity: Good<br><br>Report type: Self-report<br><br>Applicability to the Task Environment: The TEIQue-SF in particular has been amended to measure "task-related stress," which is especially applicable to the kind of stress we anticipate will influence team members in the TCT, making it a good fit for our purposes. Applicability to the Target Participant: The questions are designed to be generalizable across many populations and teaming/working conditions, which makes this a good fit for the TCT. |
| **Locus of Control** | |
| Spheres of Control (Paulhus, 1983) | Availability: Publically Available<br><br>Reliability and Validity: Good construct validity has been demonstrated for this measure using a study of athletes (Paulhus, 1983).<br><br>Report Type: Self Report<br><br>Applicability to the Task Environment: The SOC sub-scales related to the self and interpersonal control are the most applicable to the teaming research goals anticipated for the testbed. The Sociopolitical control scale is least likely to be useful.<br><br>Applicability to the Target Participant: The items in the SOC scale should be applicable to any participant population. |
| **Preference for Group Work** | |
| Preference for Group Work (Shaw, Duffy & Stark, 2000) | Availability: Publically available.<br><br>Reliability and Validity: This measure had favorable results in a psychometric analysis reported in Shaw et al., 2000).<br><br>Report Type: Self Report<br><br>Applicability to the Task Environment: The items in this measure are worded broadly and are applicable to many task environments.<br><br>Applicability to the Target Participant: The items in this measure are worded broadly and are applicable to many participant audiences. |

| Recommended Measure | Measurement Qualities |
|---|---|
| Preference for Virtual Teams (Luse, McElroy, Townsend & DeMarie 2013) | Availability: Publically available. |
| | Reliability and Validity: Items were adapted from Shaw et al. (2000) which has good established content validity. A confirmatory factor analysis on the adapted measure lends psychometric support for preference for virtual teams over working alone, and for virtual teams over face-to-face teaming. |
| | Report Type: Self-report |
| | Applicability to the Task Environment: Virtual teams are similar to distributed teams in the proposed task setting, and therefore this measure would be reasonably applicable to the task environment. However, it may be needlessly specific to the distributed aspect of teaming, rather than teaming more generally. |
| | Applicability to the Target Participant: The items of this measure are broadly worded and the measure has applicability to many participant audiences. |

The second group of constructs reviewed in this report are particular to the beliefs and attitudes that an individual has about their specific team. These constructs are more inclined to be informed by an individual's experience operating in particular group settings, and therefore possibly more likely to change with experience or training. These constructs are reviewed in Table 11.

**Table 11. Summary of Teaming Beliefs and Attitudes Recommended for Inclusion in the CONOPS**

| Recommended Measure | Measurement Qualities |
|---|---|
| **Team Efficacy** | |
| *Social Support within Team Settings*<br><br>Campion, M. A., Medsker, G. J., & Higgs, A. C. (1993). | Availability: Publically available. |
| | Reliability and Validity: This measure was predictive of positive group outcomes, but was not validated using a similar measure of social support in team settings. It is also not a generalized measure of attitudes and perceptions, but specific to one aspect of team dynamics. |
| | Report Type: Self-report |
| | Applicability to the Task Environment: The items for this measure are broadly worded and applicable to many task environments. |
| | Applicability to the Target Participant: The items for this measure are broadly worded and applicable to many participant audiences. |

| Recommended Measure | Measurement Qualities |
|---|---|
| *Communication and Cooperation within Groups*<br><br>Campion, M. A., Medsker, G. J., & Higgs, A. C. (1993). | Availability: Publically available.<br><br>Reliability and Validity: This measure was predictive of positive group outcomes, but was not validated using a similar measure of communication and cooperation in team settings. It is also not a generalized measure of attitudes and perceptions, but specific to one aspect of team dynamics.<br><br>Report Type: Self-report<br><br>Applicability to the Task Environment: The items for this measure are broadly worded and applicable to many task environments.<br><br>Applicability to the Target Participant: The items for this measure are broadly worded and applicable to many participant audiences. |
| *Individual Differences in Anthropomorphism Questionnaire (IDAQ)*<br><br>Waytz, A., Cacioppo, J., & Epley, N. (2010). Who sees human? The stability and importance of individual differences in anthropomorphism. Perspectives on *Psychological Science, 5*(3), 219-232. | Availability: Publically available.<br><br>Reliability and Validity: Złotowski et al. (2018) used the original version of the scale (30 items and an 11-point scale) and reported a Cronbach alpha of 0.92. Letheran, Kuhn, Lings, & Pope (2016) determined that a 15-item version of the original scale using a seven-point Likert scale performed well and, in fact, a nine-item version had a Cronbach alpha of 0.88.<br><br>Report Type: Self-report<br><br>Applicability to the Task Environment: The items for this measure are broadly worded and applicable to many task environments.<br><br>Applicability to the Target Participant: The items for this measure are broadly worded and applicable to many participant audiences. |
| *Self-Management*<br><br>Campion, M. A., Medsker, G. J., & Higgs, A. C. (1993). | Availability: Publically available.<br><br>Reliability and Validity: This measure was predictive of positive group outcomes, but was not validated using a similar measure of team self-management. It is also not a generalized measure of team potency, but specific to one aspect of team dynamics that may be related to potency.<br><br>Report Type: Self-report<br><br>Applicability to the Task Environment: The items for this measure are broadly worded and applicable to many task environments.<br><br>Applicability to the Target Participant: The items for this measure are broadly worded and applicable to many participant audiences. |

| Recommended Measure | Measurement Qualities |
|---|---|
| *Workload Sharing*<br><br>Campion, M. A., Medsker, G. J., & Higgs, A. C. (1993). | Availability: Publically available.<br><br>Reliability and Validity: This measure was predictive of positive group outcomes, but was not validated using any similar measure of work load. It is also not a generalized measure of team potency, but specific to one aspect of team dynamics that may be related to potency.<br><br>Report Type: Self-report<br><br>Applicability to the Task Environment: The items for this measure are broadly worded and applicable to many task environments.<br><br>Applicability to the Target Participant: The items for this measure are broadly worded and applicable to many participant audiences. |
| *Participation*<br><br>Campion, M. A., Medsker, G. J., & Higgs, A. C. (1993). | Availability: Publically available.<br><br>Reliability and Validity: This measure was predictive of positive group outcomes, but was not validated using any similar measure of inter-team participation. It is also not a generalized measure of team potency, but specific to one aspect of team dynamics that may be related to potency.<br><br>Report Type: Self-report<br><br>Applicability to the Task Environment: The items for this measure are broadly worded and applicable to many task environments.<br><br>Applicability to the Target Participant: The items for this measure are broadly worded and applicable to many participant audiences. |
| **Team Cohesion** | |
| *10-item Revised GEQ*<br><br>Carless, S. A., & De Paola, C. (2000). The measurement of cohesion in work teams. *Small Group Research, 31*(1), 71-88. | Availability: Publically available.<br><br>Reliability and Validity: Carless and DePaola's (2000) work validated their revised measure in their study (reliability was not assessed).<br><br>Report Type: Self-report<br><br>Applicability to the Task Environment: The items of this measure are reasonably broad and could support a lot of different teaming environments.<br><br>Applicability to the Target Participant: One drawback is that the questions on this measure are specific to teams that are co-located, and so would not apply to distributed teams (which may be important for the research conducted in the proposed testbed). |

| Recommended Measure | Measurement Qualities |
|---|---|
| ***Autonomous Agent Teammate-likeness (AAT)***<br><br>Wynne, K. T., & Lyons, J. B. (2018). An integrative model of autonomous agent teammate-likeness. Theoretical Issues in Ergonomics Science, 19(3), 353-374.<br><br>Wynne, K. T., & Lyons, J. B. (2019). Autonomous Agent teammate-likeness: scale development and validation. In Virtual, Augmented and Mixed Reality. Applications and Case Studies: 11th International Conference, VAMR 2019, Held as Part of the 21st HCI International Conference, HCII 2019, Orlando, FL, USA, July 26–31, 2019, Proceedings, Part II 21 (pp. 199-213). Springer International Publishing. | Availability: Publically available.<br><br>Reliability and Validity: Each of the six ATT subscales has adequate reliability (0.75 – 0.91). The dimensions were positively correlated, but multi-collinearity was absent, indicating that the dimensions captured distinct constructs<br><br>Report Type: Self-report<br><br>Applicability to the Task Environment: The items of this measure are reasonably broad and could support a lot of different teaming environments.<br><br>Applicability to the Target Participant: The measure is broadly applicable to a range of teaming environments. |